# Direct imaging elucidates ionic memory in two-dimensional nanochannels

Kalluvadi Veetil Saurav[1,2]*, Nathan Ronceray[3,9]*, Baptiste Coquinot[4], Agustin D. Pizarro[3], Ashok Keerthi[1,2,6], Theo Emmerich[3,5,#], Aleksandra Radenovic[3,7,#], Boya Radha[2,6,8,#]

*These authors contributed equally

#To whom correspondence should be addressed;
e-mail: theo.emmerich@ens-lyon.fr,
aleksandra.radenovic@epfl.ch, radha.boya@manchester.ac.uk

[1] Department of Chemistry, School of Natural Sciences, The University of Manchester, Manchester, M13 9PL, U.K.

[2] National Graphene Institute, The University of Manchester, Manchester, M13 9PL, U.K.

[3] Institute of Bioengineering, School of Engineering, EPFL, Lausanne, Switzerland

[4] Institute of Science and Technology Austria (ISTA), Am Campus 1, 3400 Klosterneuburg, Austria

[5] Laboratoire de Physique, UMR CNRS 5672, ENS de Lyon, Université de Lyon, Lyon, France

[6] Photon Science Institute, The University of Manchester, Manchester, M13 9PL, U.K.

[7] NCCR Bio-Inspired Materials, EPFL, Lausanne, Switzerland

[8] Department of Physics and Astronomy, School of Natural Sciences, The University of Manchester, Manchester, M13 9PL, U.K.

[9] Present address: Laboratoire Photonique Numérique et Nanosciences, Univ. Bordeaux, Talence, F-33400 France

## Abstract

Nanofluidic memristors promise brain-inspired information processing with ions, yet their microscopic origin remains debated. So far, ionic memory has been attributed to ion-specific interactions, dynamic wetting, chemical reactions or mechanical deformations, yet typically without direct experimental evidence. Here, by combining operando interferometric imaging with electrokinetic measurements, we directly visualize voltage-induced blistering of the confining walls of two-dimensional (2D) nanochannels, as key origin of memristive hysteresis. We identify two distinct classes of blisters: unidirectional, driven by electrostatic forces on surface charges, and bidirectional, arising from osmotic pressure due to concentration polarization. This mechanistic framework explains device evolution and device-to-device variability, and reframes stochastic blistering as a functional design element. Our results constitute a direct proof of electromechanical coupling as a robust pathway to ionic memory in 2D nanochannels and open routes toward high-performance ionic memristors and electrically actuated nanofluidic valves.

## Main text

Emerging phenomena in nanoscale ion transport under extreme confinement are a major interest of nanofluidics[1,2], both for the fundamental questions they raise and for their potential to enable novel applications. In particular, striking nonlinearities such as hyperbolic currents[3], diode-like rectification[4–8] and memristive hysteresis[9–14] have been experimentally observed in a variety of nanofluidic devices. Memristive nanofluidic devices have attracted significant attention by enabling neuromorphic nanofluidics, a new field aiming at processing information with solvated ions, thereby mimicking the brain[15–17]. In most studies, the physical origin of nonlinear ion transport is rationalized either through molecular simulations or through phenomenological models involving *ad hoc* new physics in extreme confinement. By relying on idealized devices, these approaches often fail at capturing the experimentally observed devices' stochasticity and evolution, which is a major obstacle toward the realization of reliable nanopore-based technologies[18–20]. Thus, operando imaging approaches are much needed to complement “blind” electrokinetic measurements and uncover unambiguously the mechanisms of nanoscale ion transport.

A recent study combined optical and electrokinetic measurements to reveal the origin of ionic memory in highly asymmetric channels, showing that voltage-induced mechanical deformations modulate the confinement and, consequently, the conductance of the device[21]. This resonates with biology, as biological ion channels, such as voltage-controlled synaptic calcium channels, also undergo mechanical deformations to regulate the transmission of information through neurotransmitter release[22]. Such results also demonstrate the potential of brain-inspired nanofluidics for phenomenologically accurate neuromorphic computing using only salt and water as ingredients. However, the working principle of many other nanofluidic memristors remains an open question. In two-dimensional (2D) nanochannels made by van der Waals assembly for instance, it has been hypothesized that ion-specific effects such as ionic pairing and polyelectrolyte formation were at the root of ionic memory[23,13,24,25]. Yet, a direct experimental proof is still lacking. Here, we combine optical thin film interferometry and electrokinetic measurements to directly observe the phenomenon triggering memristive hysteresis in two-dimensional nanochannels. We observe that voltage-induced mechanical deformations lead to nonlinear ion transport in 2D channels. We further identify the electromechanical effects that explain these deformations through systematic characterization and analytical modelling.

### Design and correlative measurements of 2D nanochannels

For the fabrication of a 2D nanochannel device, a modified version of the fabrication process (Fig. S1) previously reported was followed[26,27], wherein a bottom hexagonal boron nitride (hBN) flake is transferred onto a silicon nitride ($SiN_x$) free-standing membrane with a micro-hole of ~ 3×25 µm. A few layer graphite flake was then patterned using electron beam lithography, followed by dry etching to be used as a spacer layer consisting in an array of parallel line-shaped strips that serve as nanochannels. Figure S1c shows atomic force microscopy (AFM) image of the spacer. The height of the spacer from the AFM image is $h$ ~ 1.7 nm which corresponds to 5-layer graphene and the channel width is ~100-110 nm and each of the channels is separated by ~1 µm (details of the fabrication steps in supplementary section 1). To confine the fluid, another top hBN flake was transferred onto the spacer flake, which was then transferred together onto the back-etched bottom flake on $SiN_x$ membrane to achieve a top-spacer-bottom trilayer stack. To define the channel length, a rectangular aperture was patterned on to the trilayer stack by photolithography at a specific distance ($L \approx 5$ µm), followed by dry etching (Fig. 1a). Importantly, the top and bottom thickness $t_{top}$ and $t_{bot}$ changed not only the mechanical but also the optical properties of the devices and had to be chosen carefully. The full details on the $N$=14 devices measured in this study are provided in Supplementary Table 1.

Our measurement setup consists of a custom-made fluidic cell with a top reservoir large enough to fit a water dipping objective (Fig. 1b). The device is observed *via* wide-field reflection imaging with quasi-monochromatic normal incidence illumination enabling the quantification of the device topography by using thin film interferometry[21,28]. Ion transport is induced by applying a sinusoidal potential drop $\Delta V(t) = \Delta V_o \sin(2\pi f t)$ with varying amplitude $\Delta V_o$ and frequency $f$ across Ag/AgCl electrodes connected to a digital/analog converter (DAC) for recording. The camera is triggered with the DAC in order to synchronize optical and electrokinetic data (Fig. S2). In this study, we used aqueous solution of potassium chloride (KCl) in various concentrations from 1 mM to 1 M as well as 1 M solutions of sodium chloride (NaCl), lithium chloride (LiCl) and calcium chloride ($CaCl_2$) (Fig. S6). Our coupled electrokinetic–optical approach is motivated by the striking diversity of current–voltage characteristics previously reported in nominally identical 2D nanochannels[13,29,30]. As illustrated in Figure 1c, these include linear but overconducting traces, strongly rectifying behaviours, and memristive hysteresis that can be either unidirectional or bidirectional. Such variability is difficult to reconcile with purely electrokinetic models, leading us to hypothesize that it instead originates from voltage-induced mechanical deformations that are naturally diverse in their manifestation[21].

**Quantifying nanoscale deformations *via* image voltage-induced contrast (VIC)**

We found indeed that the application of a voltage resulted in visible features, namely liquid-filled blisters of nanoscale height $h_B$ and micron-scale lateral extension. In this work, we introduce a methodology for the quantification of the voltage-induced deformations of the device (Fig. 1d) through the voltage-induced contrast (VIC) $\Psi(\Delta V)$ defined for each pixel by:

$$\Psi(\Delta V) = \frac{\eta(\Delta V) - \eta(0)}{\eta(\Delta V) + \eta(0)} \quad (1)$$

where $\eta(\Delta V)$ is the pixel image intensity value for an applied voltage ΔV. This contrast metric can be understood as the voltage-induced Michelson contrast and has values between -1 and +1. Its value depends on the thickness of all the layers traversed by the light, where each interface contributes to reflection and propagation through each layer introduces a dephasing. We implemented a full transfer matrix method (TMM) calculation of the device contrast in the geometries used in this work (details in Supplementary section 5). For a given device, the only parameter susceptible to evolving under the application of a voltage is the blister height $h_B$. Our TMM simulations showed that for the device parameters used in this work, the contrast consistently increases from 0 to positive values upon blister formation (Fig. 1e). The layered structure was determined by optical contrast on the exfoliation wafer, with thicknesses known within a ~10 nm wide window around target values, resulting in small contrast changes highlighted in Figure 1e.

For small deformations, the VIC $\Psi$ corresponding to a blister of height $h_B(\Delta V)$ reads:

$$\Psi(\Delta V) = 4\pi\, n_w\, h_B(\Delta V) / \lambda \cdot F(n_j, t_j, \lambda) \tag{2}$$

where $n_w$ is the refractive index of water and $F(n_j, t_j, \lambda)$ is a dimensionless parameter of order 1 which depends on the layer thicknesses $t_j$ and refractive indices $n_j$ of the layers $j = 0,...,5$. $F(n_j, t_j, \lambda)$ defines the deformation sensitivity of the experiment, and we chose $\lambda$ = 635 nm to maximize it for the relevant device parameters. We detail in the supplementary information S5 the full TMM contrast calculation for the 6-layer thin film heterostructure. We quantified the impact of layer thickness uncertainty derived from the ~10 nm window of our exfoliation metrology, which results in an uncertainty $\Delta h_B/h_B = \Delta F/F \sim 30\%$ for contrast-based height estimates. This bound ensures that the contrast remains monotonic for small deformation magnitudes across all devices (Fig. S3-S5). Thus, we primarily report the model-independent contrast Ψ throughout this work, treating converted heights as order-of-magnitude estimates.

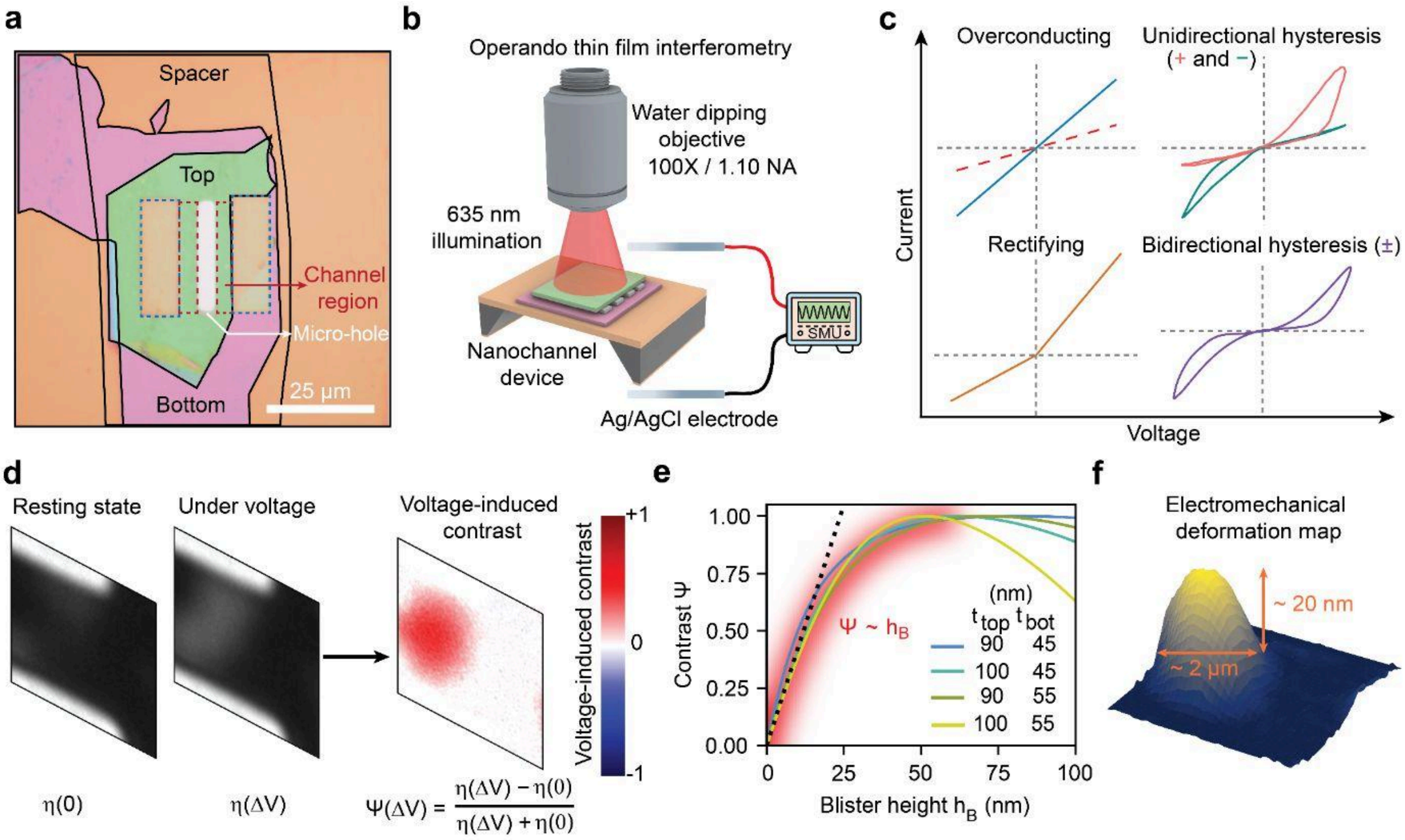

**Figure 1. Operando microscopy of a 2D nanochannel device and methodology. a,** Optical image of device 2 (5-layer graphene spacer) depicting the trilayer stack of hBN-Gr-hBN on $SiN_x$ membrane with a ~3×25 µm micro-hole and rectangular aperture separated from the micro-hole by a fixed length defined as channel length $L$. **b,** Schematic of operando optofluidic measurement setup wherein a water-dipping objective is used to visualize the dynamics of 2D nanochannel while simultaneously recording the ionic current. **c,** Representative $I$–$V$ characteristics namely over-conducting (the dashed line represents theoretical current value), rectifying and two types of memristors (unidirectional hysteresis and bidirectional hysteresis) as observed and discussed in the paper. **d,** Image processing methodology for the quantification of voltage-induced blisters from raw reflectance images. The VIC $\Psi(\Delta V)$ is evaluated as the Michelson contrast of the raw intensity images at rest η(0) and under an applied voltage η($\Delta V$). **e,** Contrast-height curves $\Psi(h_B)$ for various top and bottom layer thicknesses ($t_{top}$ and $t_{bot}$ in nm) used in this study. The highlighted red region shows the range of monotonous scaling between $\Psi(h_B)$ and $h_B$. The dotted line corresponds to equation (2) valid for small deformations. **f,** Blister topography extracted from **d-e**.

**Ion transport nonlinearities originate from voltage-driven blister formation**

We measured a variety of nonlinear *I-V* characteristics, in accordance with previous reports in similar devices[13,29], which we investigated with our correlative approach. Varying the sinusoidal voltage amplitude $\Delta V_o$ at high salt concentration (1 M KCl) from 0.05 V to 1 V (Fig. S7), we observed linear ion conduction at low voltage amplitude (Fig. 2a, $\Delta V_o<V_c \approx 0.2$ V). Although the *I-V* characteristic is linear, the measured conductance is higher than the expected theoretical value by several fold (Fig. 1c). This can be understood from the raw intensity image presented in Figure 2b1, where areas with different contrast are circled in yellow, identifying liquid-filled blisters. This previously overlooked imperfect adhesion could result in an underestimation of the effective channel cross-section in addition to the previously proposed contribution of interfacial hydroxide ions[30]. In this regime of linear conduction, no significant VIC is observed at either positive or negative voltages (Fig. 2b2-3). In other words, the blisters circled in yellow in Figure 2b1 are not voltage-responsive, as sketched in the top-left inset of Figure 2a. For larger applied voltages $\Delta V_o>V_c$, we observed the onset of nonlinearities including memristive hysteresis (Fig. 2d), coinciding with the appearance of blisters at high absolute value of the voltage (green circles, Fig. 2e). Blister formation and/or enlargement inside the nanochannel area have the effect of temporarily increasing the device's cross-section, thereby increasing its conductance. This shows the electromechanical origin of the nonlinearity of the *I-V* characteristic. Crucially, the strictly linear response observed in the absence of dynamic blister deformation (Fig. 2a) demonstrates that active structural evolution is required for hysteresis (see Supplementary Discussion).

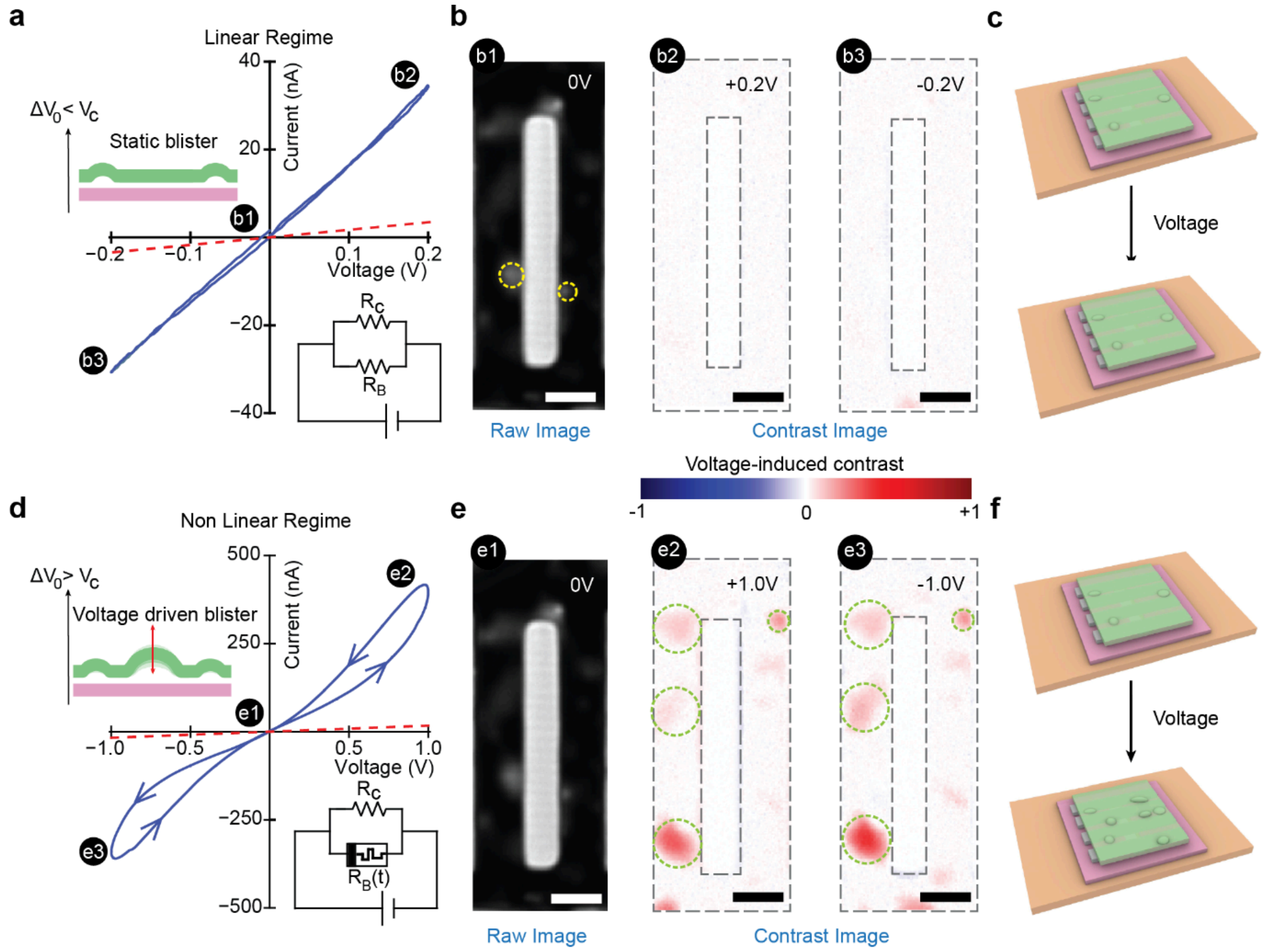

**Figure 2. Origin of memory in 2D nanochannels.** *I-V* characteristic of device 1 (5-layer graphene spacer) in 1 M KCl for a driving voltage $\Delta V(t)=\Delta V_o \sin(2\pi f t)$ at a frequency $f$ = 100mHz showing transition from linear ion transport phenomenon to memristive hysteresis. **a,** *I-V* characteristic for $\Delta V_o$= 0.2 V showing a linear ion transport regime with no hysteretic behaviour in the curve. **b,** Corresponding snapshot of the raw image at 0V (b1) and VIC images (after post processing) captured at +0.2 V (b2), and -0.2 V (b3) respectively, shows no evidence of deformations or blister formation within the nanochannel. **c,** Schematic image of the evolution of blisters in a nanochannel device under applied voltage $\Delta V_o$= 0.2 V**. d,** *I-V* characteristic for $\Delta V_o$= 1 V showing a nonlinear ion transport regime with hysteretic behaviour. Directional arrows indicate the hysteresis loop direction. **e,** Corresponding snapshot of raw image at 0 V (e1) and VIC images show evidence of deformations (circled in green) at +1 V (e2) and -1 V (e3), respectively. **f,** Schematic image of the evolution of blisters in a nanochannel device under applied voltage $\Delta V_o$= 1 V**.** The top-left inset in both *I–V* curves (2a and d) shows a schematic of the device structure with blisters, corresponding to the conduction behaviour, while the bottom-right inset depicts the equivalent circuit where $R_C$ and $R_B$ are the channels and the blister resistances, respectively. In the nonlinear case, $R_B(t)$ is a memristor. The dashed red lines in both the *I-V* graphs represent the theoretical current obtained from the theoretical conductance, calculated based on the device parameters. Scale bar: 5 µm. Supplementary video 1 presents the full data in **d-e**.

## Blister stochasticity explains ion transport anomalies and evolution

We applied our correlative approach to devices exhibiting *I-V* characteristics of different shapes, to investigate their distinctive features and localize the device regions responsible for the nonlinear response. In Figure 3a, we show measurements of device 2 over 4 cycles, with a clear and repeatable conductance jump around -0.7 V (conductance +50% over <1 s). To localize the source of this jump in the complex patterns arising in the VIC images, we analyzed kymographs of the VIC along the left and right yellow lines along the channel sections (Fig. 3b). The kymographs exhibit several periodic patterns with substantial dephasing, but a clear VIC change can be found in the right channel region blister 2l, exactly matching the conductance jump apparent in the *I-V* characteristic, as shown by the brown arrows. This directly shows the link between fast and local change of confinement and overall device conductance. Further, it explains conductance jump effects in 2D slits which are not captured by ion-specific phenomenological modeling. Additionally, the top-left blister 2m has a negative VIC for the part that is not covered by the top layer, in accordance with our TMM calculations (Fig. S5). This demonstrates the possibility of blister formation between the bottom layer and the $SiN_x$ membrane (Fig. S8a), which likely contributes a portion of the observed conductance changes together with blister 2n below.

We found occurrences of devices exhibiting an irreversible evolution towards either higher or lower conductance states, and we examined this scenario in device 6, presented in Figure 3c where the 4 cycles lead to non-overlapping *I-V* characteristics. As a side note, this device displays considerable conductance jumps resulting in a large ON/OFF ratio of 50, suggesting a large potential of 2D slits and associated blistering effects for neuromorphic computing. The VIC images revealed two voltage-sensitive regions ROI 1 and ROI 2 corresponding to the formation of blisters with a negative VIC. As shown in Figure S4, negative VIC upon blister formation is expected for this particular device geometry (used in previous studies[13,14,29]). Note that the irreversible changes reported here were also observed in devices without this gold patch, and conversely gold patch devices were not all prone to irreversible evolution in similar experimental conditions. The kymograph analysis of device 6 in Figure 3d shows that while the conductance time trace is complex and stochastic, its complex response can be broken down to the contributions of the two blisters of ROI 1 and ROI 2 which account for all the conductance evolution of the device along the cycles. In particular, when the blister in ROI 1 skips the 4$^{th}$ cycle, the conductance does not increase as for the first three cycles in the corresponding time window. Likewise, the abrupt conductance jumps identified with the brown arrows coincide precisely with abrupt VIC jumps of ROI 2. For gold patch devices (Fig. 3c-d and Fig. S8c), we found VIC values not exceeding 0.1 in absolute value, which indicates deformations under ~5 nm (Fig. S4).

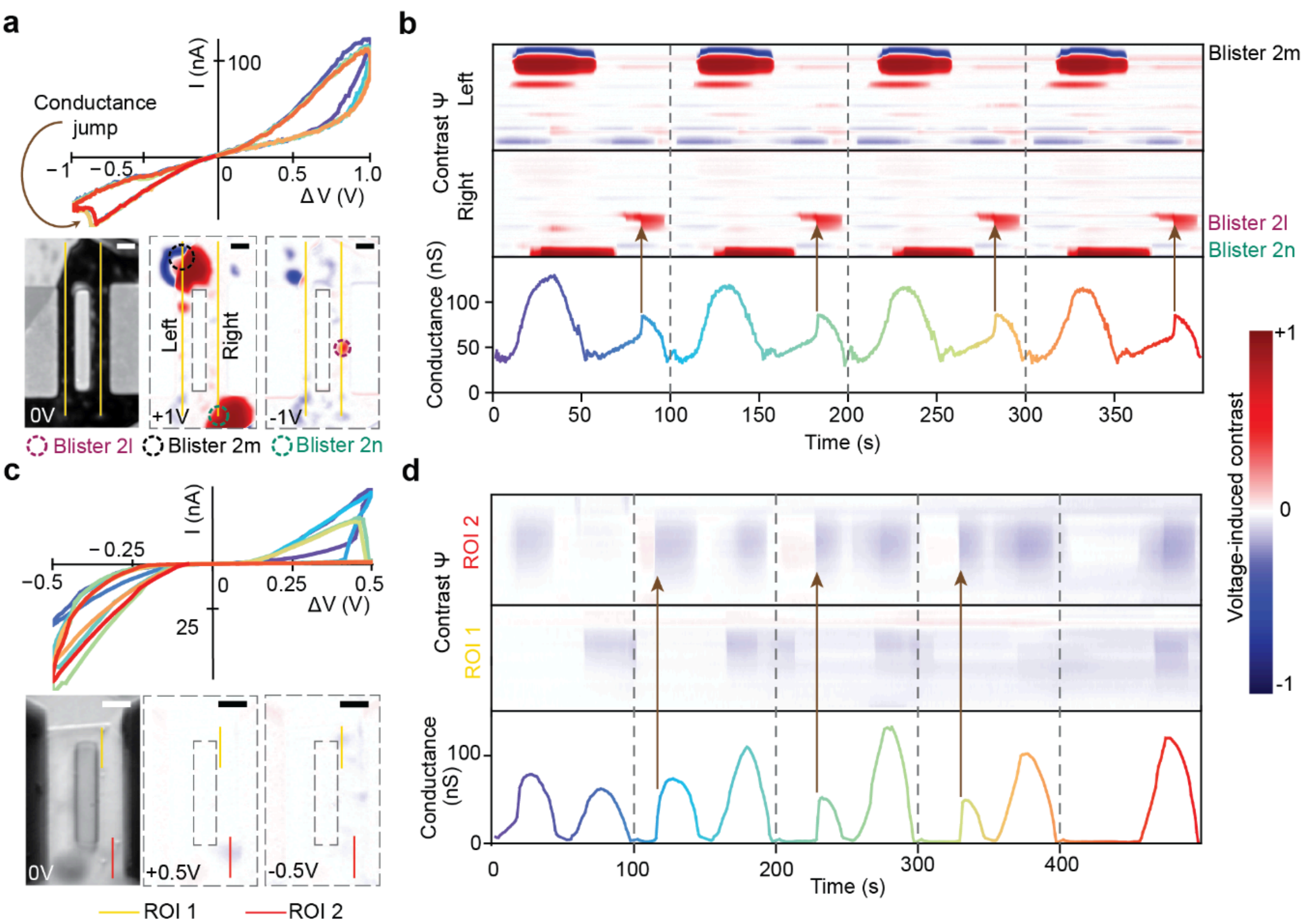

**Figure 3. Spatiotemporal analysis of conductance jumps and device stochastic evolution. a,** Repeated *I-V* characteristics of device 2 (1 M KCl at a frequency *f* =10 mHz) exhibiting a conductance jump around -0.7 V, and corresponding raw image at 0 V and VIC images at ±1 V. **b,** VIC kymographs extracted along the lines labelled 'Left' and 'Right' in **a**, correlated with the conductance time trace. While many voltage-responsive blisters are present, this analysis enables pinpointing which blister opens the ionic path creating the conductance jump effect. **c,** Consecutive *I-V* characteristics and corresponding images of the first cycle for device 6, which includes a gold patch. The device exhibited hysteresis that evolved over cycles as shown by the non-overlapping characteristics. **d,** VIC kymographs extracted along the yellow and red lines in **c**, showing the two voltage-responsive regions of the device. Both react to voltage in a stochastic manner, and together ROI 1 and ROI 2 account for all conductance variations observed in the time trace. The time color codes of the *I-V* characteristics match the respective conductance time traces. The dashed lines in panels **b** and **d** indicate individual voltage cycles. Scale bar: 5 µm.

By comparing the two aforementioned devices, we conclude that a highly repeatable *I-V* curve is the consequence of a stable blister landscape. The observed voltage-driven blistering phenomenology revealed through our correlative measurement approach thus explains the conduction anomalies in the devices 1, 2, 3 and the other 11 devices reported in this work.

To further characterize these deformations, we next examined how individual blisters respond to the polarity of the applied voltage. This directionality analysis revealed two distinct response types: blisters that expand at both voltage polarities (Device 1, Fig. 4a, shown in Supplementary video 1) and blisters that expand only under one polarity (Device 2, Fig. 4b, shown in Supplementary video 2). This distinction already points to different underlying driving forces, which we explore more quantitatively in the following sections (corresponding *I*-*V* in Fig. S9). The VIC time traces of the blisters circled in Figure 4a are presented in Figure 4c. We find that just like the conductance time trace, all three VIC traces consistently increase at both positive and negative voltages. On the contrary, for device 2 the VIC and conductance increases occur only at positive voltages while no blister forms at negative voltages (Fig. 4d). This directionality analysis can be visualized through contrast-voltage ($\Psi$-$V$) characteristic: for device 1, all three blisters had a near-symmetrical bidirectional $\Psi$-$V$ characteristics (Fig. 4e), while all three blisters of device 2 had a unidirectional positive $\Psi$-$V$ characteristic (Fig. 4f). Figure S10 shows the conductance vs VIC for bidirectional (device 1) and unidirectional (device 2) hysteresis, respectively. Additionally, unidirectional blisters of Figure 4f exhibit different threshold voltages for blister formation: blister 2i forms at near-zero voltage, while blister 2j and blister 2k form around +0.25 V. Once this voltage threshold is overcome, the blister VIC evolves progressively to a maximum value that corresponds to the maximum blister height on the order of 5 to 25 nm according to Figure 1e.

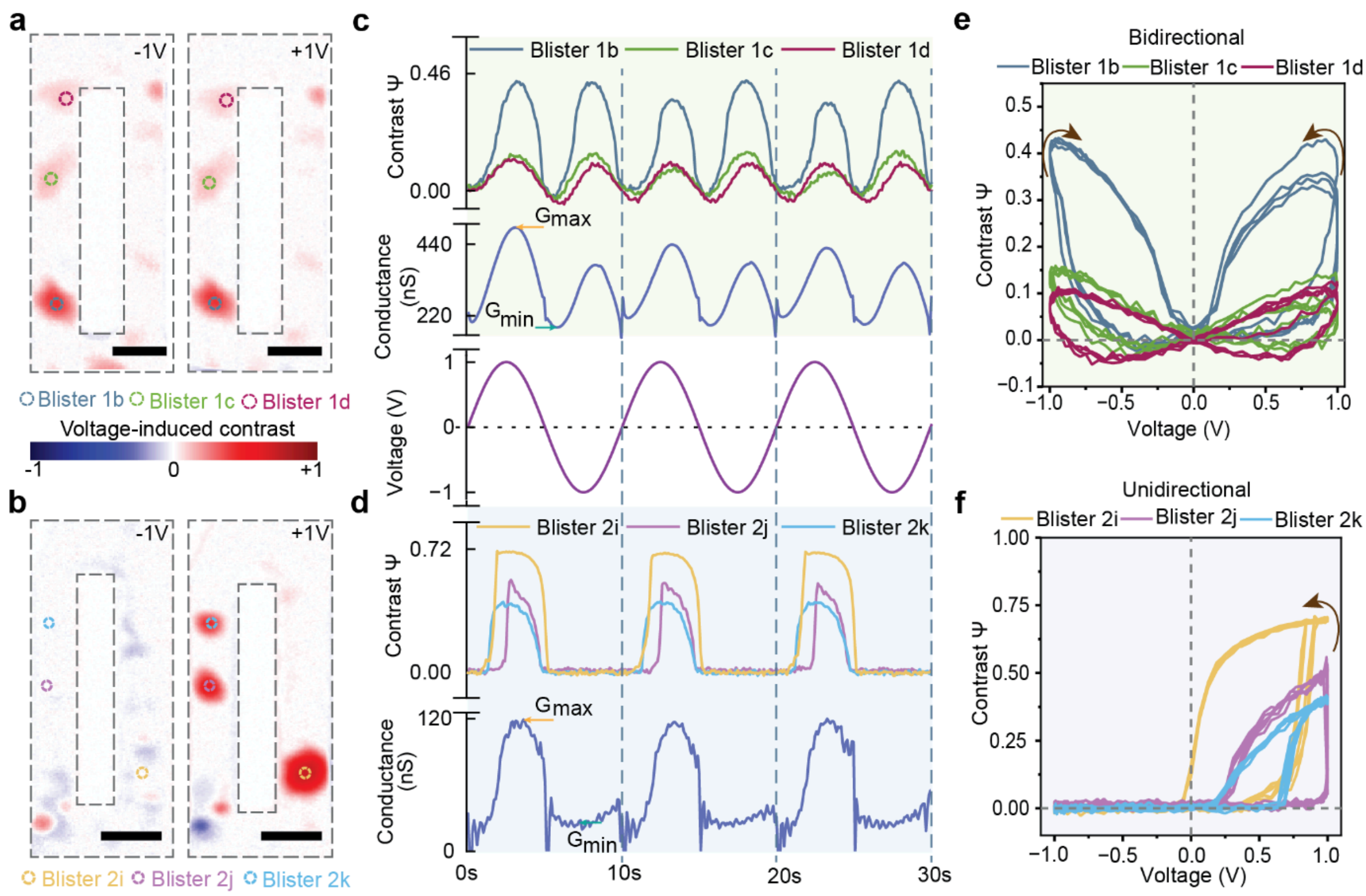

**Figure 4. Imaging reveals two types of voltage-responsive blisters. a,** VIC images of device 1 at ± 1V, showing near-symmetric deformations for positive and negative voltage. **b,** VIC images of device 2 at ± 1V, showing unidirectional deformations for positive voltage only. **c,** VIC time traces of device 1 taken at the ROIs highlighted in **a** and the conductance trace increase regardless of the sign of the voltage forcing. **d,** VIC time traces of device 2 taken at the ROIs highlighted in **b** and the conductance trace increases only at positive voltages.**e**, The $\Psi$-$V$ characteristics of the ROIs show a near-symmetric bidirectional pattern. **f,** The $\Psi$-$V$ characteristics of the ROIs show a unidirectional positive pattern. The dashed vertical lines in panels **c** and **d** indicate individual voltage cycles. Scale bar: 5 µm.

While these two devices exhibit uniform blister directionality, we also observed hybrid cases featuring a mixture of bidirectional and unidirectional blisters (Fig. S6, S17), which yield complex asymmetrical *I-V* characteristics. This phenomenological diversity extends to the dynamics and retention of these electromechanical deformations. Frequency-dependent measurements revealed that both the macroscopic *I-V* hysteresis and individual blister loop areas decrease with increasing driving frequency (Figs. S12-S15). Crucially, by monitoring the structural state of the channels at the zero-voltage crossing during these sweeps, we uncovered the physical origin of memory retention. While the majority of deformations are purely volatile and collapse fully at 0 V, particularly at higher driving frequencies (Fig. S17), a specific subset of blisters sustains a non-zero structural deformation even after the bias is removed (Fig. S16). These non-volatile states emerge primarily at lower frequencies for device 2 (10–30 mHz) where unidirectional blisters have sufficient time to grow and reach large-scale deformations required for memory retention (Fig. S17a-b).

To contextualize these individual dynamic observations and gain a quantitative picture of the electromechanical landscape underlying ionic memory, we systematically analyzed the voltage-induced blister dynamics across multiple devices (*N=10*) and compared them with transport. Figure 5a illustrates the extraction of two descriptors from the VIC $\Psi(t)$ and the normalized conductance variation $\delta G(t) = (G(t)-G_{min})/(G_{max}-G_{min})$: the symmetry ratio $S = A_{small}/A_{large}$, which measures whether blister expansion is bidirectional ($S \sim 1$) or unidirectional ($S \ll 1$), and the contrast–conductance correlation $C = \langle \delta G(t) \times \Psi(t) \rangle$, which quantifies how strongly blister VIC co-varies with normalized conductance. Applying these metrics to all blisters ($N$ = 47 across 10 devices), we find that the symmetry ratio clusters cleanly into two groups: unidirectional blisters with low S and bidirectional blisters with S near unity. By contrast, the correlation *C* displays a broader spread, reflecting the fact that the impact of a single blister on conductance depends not only on its underlying mechanism, but also on its location within the channel and on the presence of other active blisters in the same device. High values of *C* nevertheless identify cases where blister dynamics dominate conductance, consolidating the causal link between local electromechanical activity and memristive ion transport. The spatial map of blister positions across devices (Fig. 5c) further reveals their stochastic nature: blisters nucleate unpredictably throughout the active region without preferred sites. Regarding the preferential formation sites of blisters, bidirectional are consistently found on sites contributing to the ion path, whereas this is not strictly required for unidirectional ones (Fig. S23c). This explains the strong device-to-device variability and temporal evolution (Fig. S18-20) in nominally identical channels, and highlights stochastic blistering as the unifying origin of the diverse memristive behaviours observed in 2D nanofluidics.

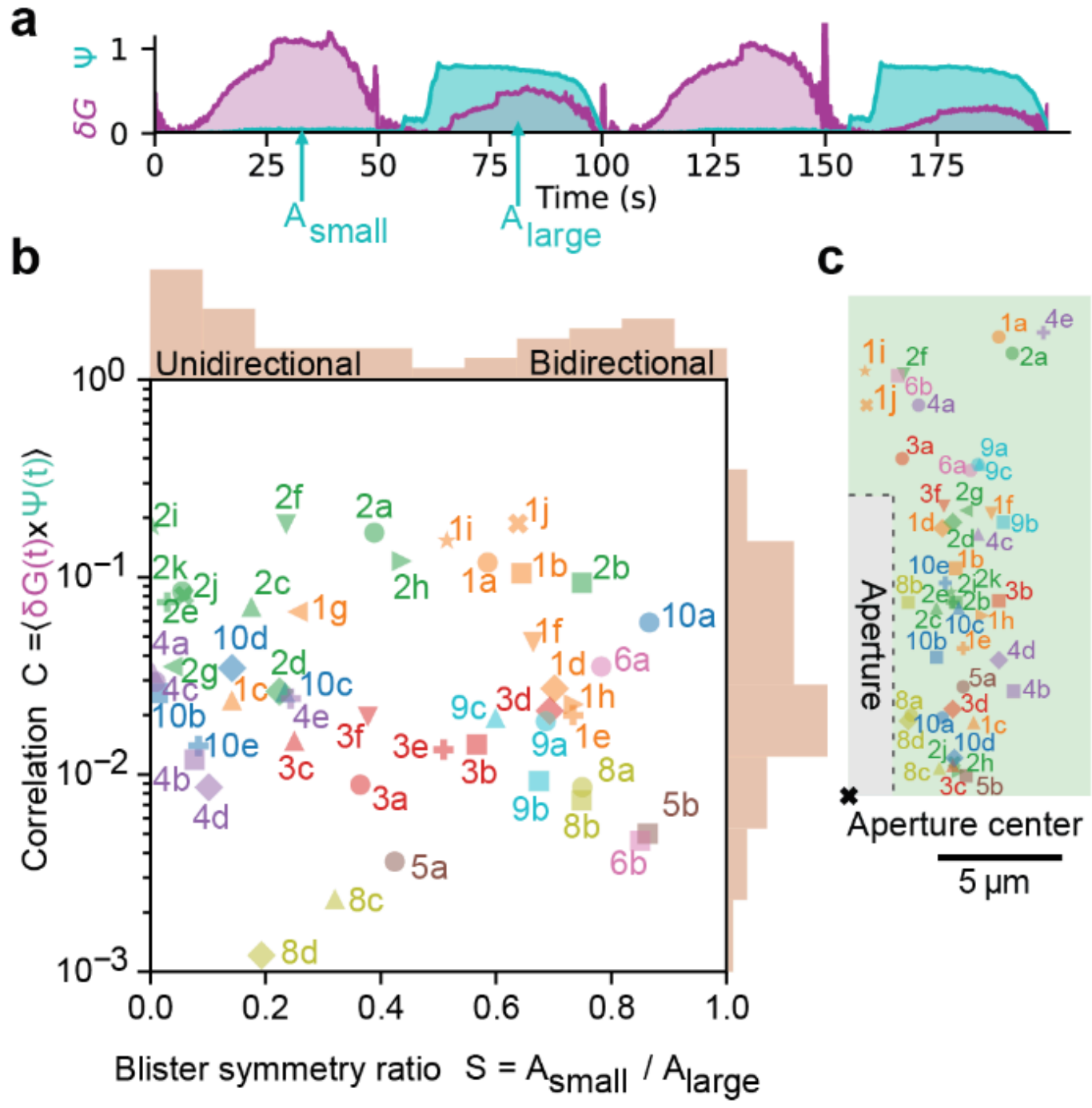

**Figure 5. Statistical analysis of blister dynamics and correlation with ion transport.**
**a**, Example trace of normalized conductance change $\delta G$ (magenta) and VIC $\Psi$ (cyan) for blister 2g, illustrating the definition of the blister symmetry ratio $S=A_{small}/A_{large}$ and the time-averaged overlap between VIC and conductance $C=\langle \delta G \times \Psi \rangle$. **b**, Scatter plot of $C$ versus $S$ for all blisters analyzed in hBN devices (N = 47 blisters from 10 devices). Two populations emerge: unidirectional ($S \ll 1$) and bidirectional ($S$~1). **c**, Spatial distribution of blister positions relative to the aperture center, overlaid for all devices. All locations were collapsed into one quadrant owing to the device's double axial symmetry. Blister coordinates are given relative to the black cross at the bottom-left, indicating the aperture center. See Fig. S23 for the full *C-S* plot with error bars and exhaustive blister positions.

## Theoretical modelling of the observed deformations

Having demonstrated that the ion conduction nonlinearities originate from blister (de)formation in between the layers of the 2D nanochannels, we turned to physical modeling of this system to identify the source of these electromechanical effects. We first estimated the pressure $P_c$ required to induce the observed deformations. For this, we derived the full energy balance of blister formation[31] including the contributions of the adhesion energy $\Gamma$ of the hBN crystal on the spacer, the elastic energy of the deformed layer, dominated by the stretching due to the layered structure of hBN,[32] and the work of the pressure force as depicted in Fig. 6a (calculation details in supplementary material section 9.1). We find that larger blisters are energetically favourable, yet in practice the size of the blisters is limited by the size of the channel region, yielding the experimentally observed blisters of radius of the order of $R \approx$ 1-2.5 µm. The corresponding detachment pressure $P_c$ required to cross the adhesion energy barrier and to generate a blister is:

$$P_c \approx \frac{(2Y_{\parallel} t)^{1/4} (8\Gamma)^{3/4}}{3R} \tag{3}$$

Where $Y_{//}$ is the in-plane Young modulus of hBN and $t$ is the thickness of the deformed layer. Note that while we cannot be sure whether the blister appears on the top or the bottom layer in devices with both hBN top and bottom walls, we could verify the occurrence of blisters forming both between the top and bottom layer (device 1-4) and between the $SiN_x$ substrate and the bottom layer (device 9, 10 and 14) by using an opaque graphite layer as the bottom or top layer, respectively (Fig.

S21). Once this threshold has been crossed ($P>P_c$), there is no more adhesion and the blister's height reads:

$$h_B \approx \left(\frac{3PR^4}{2Y_{\parallel}t}\right)^{1/3} \tag{4}$$

According to this formula, the experimental data correspond to pressure forces of the order of 1 bar, which is typical of nanofluidic phenomena[33,34]. However, for the detachment pressure $P_c$ to be attainable, we found that the adhesion energy $\Gamma$ must be two orders of magnitude lower than its maximal value on pristine hBN/graphene interfaces[35] ($\Gamma_{max}$ = 150 mJ/m$^2$). This reduced adhesion could be due to the persistence of nanometer thin hydrocarbon films on 2D materials surfaces even after annealing, which are known to significantly lower interfacial adhesion energies and may be remodelled in fluidic environments[36–38]. The adhesion is nevertheless crucial since it may explain the memory effect as sketched in Figure 6b. Indeed, while the adhesion barrier must be overpassed by the growing blister, it does not apply on the receding blister, which then simply follows equation (4). Thus, we predict that the blister height, and thus the conductance, exhibits hysteresis during a voltage cycle. In practice, we also expect the adhesion landscape to be disordered, resulting in a smoothed hysteresis. Dynamically, we expect the filling of the growing blister, once the threshold has been overcome, to be limitant, leading to a smoothened hysteresis. We further predict this hydrodynamic effect to suppress the memory at frequencies large compared to 100 mHz as the blister has no longer the time to grow (see supplementary material section 9.1.5). Regarding the source of the detachment pressure, we found two phenomena which can provide the required pressure and either a bidirectional voltage response or a unidirectional one.

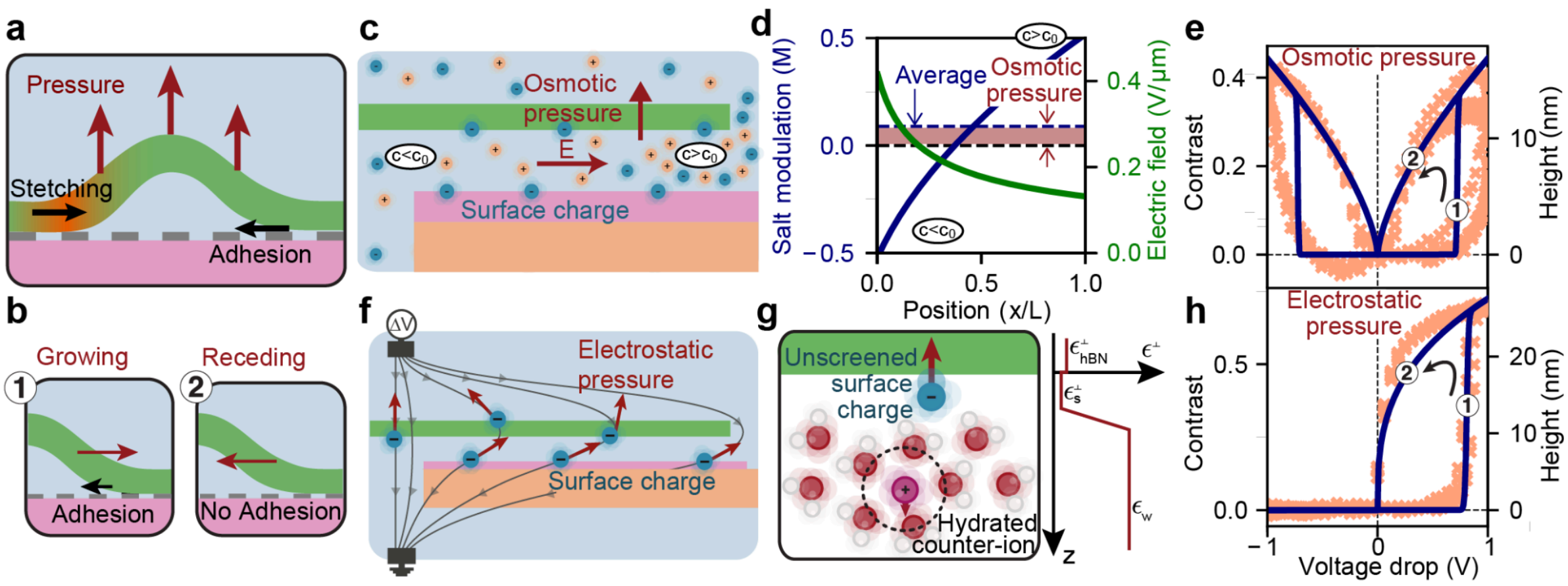

**Figure 6**: **Mechanism of voltage-induced blister (de)formation. a,** Sketch of a blister, pulled by an upward pressure and limited by the stretch and the adhesion on the spacers. **b,** Sketch of the growing and receding of a blister. The presence of the adhesion energy barrier while the blisters grow but not when it decreases, induces a mechanical hysteresis leading to ion transport memory. **c,** Sketch of the nanochannel in the presence of an electric field: the transport of the excess of cations due to the surface charge generates a concentration polarisation along the channel. **d,** In blue, the salt concentration modulation compared to the equilibrium concentration, $c(x)$-$c_o$-$|\Sigma|/h$ along the channel. The dashed line is the average salt concentration excess, which leads to the osmotic pressure. In green, the corresponding electric field along the channel, modulated by the local ionic resistivity. **e,** Prediction of blister height and the corresponding contrast as a function of the voltage in a cycle. The arrow indicates the direction of the cycle and the numbers refer to panel b. The orange crosses correspond to the experimental data of the blister 1b (see Fig. 4) for which we used an effective radius of R = 1 µm and fitted an effective adhesion $\Gamma$ = 1.5 mJ/m$^2$ and a surface charge of $\Sigma$ = -16 mC/m$^2$. **f,**

Sketch of the device with electric field lines crossing the top and bottom layers and pulling surface charges. **g,** Zoom on a surface charge and its counter-charge. While the counter-charge is surrounded by water molecules efficiently screening the ion ($\epsilon_w$), the surface charge is only weakly screened in the normal direction ($\epsilon^{\perp}_s \ll \epsilon_w$), resulting in a stronger electrostatic force acting on surface charges than on counter ions. **h,** Prediction of blister height and the corresponding contrast as a function of the voltage in a cycle. The arrow indicates the direction of the cycle and the numbers refer to panel b. The orange crosses correspond to the experimental data of the blister 2i (see Fig. 4) for which we used an effective radius of R = 1.5 µm and fitted an effective adhesion $\Gamma$ = 4.1 mJ/m$^2$ and a surface charge of Σ = -22 mC/m$^2$.

In the first model, we considered the effect of the in-plane electric field along the channel of uniform negative surface charge Σ (Fig. 6c). While the Dukhin number $Du = |\Sigma|/ec_o h \approx 0.2$ remains modest at 1 M it increases at smaller concentrations, and its coupling with the strong electric potential drop $e\Delta V/k_B T \approx 40$ for $\Delta V$ = 1 V is sufficient to generate a significant concentration polarisation (Supplementary Section 9.2)[39]. Indeed, the larger concentration of cations in the channel dragged by the electric field will result in an over-concentration of salt on one side of the channel, and a depletion on the other side. Since the channel length $L \approx 5$ µm is much larger than the Debye length $\lambda_D = (\epsilon k_B T/2e^2 c_o)^{1/2} \approx 1$ nm, we assumed electroneutrality and that the local polarisation only accommodates the electric field to the local ionic resistivity, which scales with $1/c(x)$. As a minimal model, we further assumed symmetric reservoirs and a strong polarisation effect to deduce the salt concentration profile (Fig. 6d) whose average is always larger than the equilibrium concentration. We ultimately found that this extra salt concentration generates an average osmotic pressure force between the inner channel and the reservoir, pushing the top layer upward:

$$P_{\mathrm{osm}} = \frac{Du}{1+Du} \frac{|\Sigma|\, e\, \Delta V^2}{96\, h\, k_B T} \tag{5}$$

With surface charge values[29,40,41] on the order of -30 mC/m$^2$, an average channel height $h \approx 3$ nm and a bulk salt concentration $c_o$=1 M, this predicts an osmotic pressure of up to 4 bar. In this model, we obtained an overpressure in the channel regardless of the sign of the voltage drop $\Delta V$, leading to bidirectional memristors whose height over a voltage cycle is plotted in Figure 6e and compares well with experimentally observed blisters (in orange). Injecting Eq. (5) into Eq. (4) further predicts that $h_B \propto \Delta V^{2/3}$, once the adhesion has been overcome, which is consistent with experimental data (see Fig. S24a).

In the second model, we considered the effect of the out-of-plane electric field ($E_\perp$). Indeed, due to the geometry of the nanofluidic device, not only the electric field lines go along the channels, but they also cross the top and bottom layers, as sketched in Figure 6f. Thus, the surface charges at the interfaces are pulled in the normal direction, for which they are weakly screened ($\epsilon^{\perp}_s \approx 2.1$), on the contrary to their counter charges surrounded by water molecules ($\epsilon_w \approx 80$) as sketched in Figure 6g[42]. Under an applied voltage, the surface charge Σ then experiences the local electric field, which induces an electrostatic pressure on the wall (Supplementary section 9.3):

$$P_{\mathrm{elec}} = 2\Sigma E_\perp \sim -\frac{2\Sigma\Delta V}{t\,\epsilon^{\perp}_s} \tag{6}$$

With $\Sigma$=-30 mC/m$^2$, we predicted electrostatic pressures up to 3 bar. In this model, the direction of the pressure force depends on the sign of the voltage drop $\Delta V$. To generate a blister on the top layer, the pressure must be upward, corresponding to a positive voltage drop $\Delta V$ in the case of a negative surface charge. As shown in Figure 6h, we then expect to form positive unidirectional memristors with the electrostatic pressure, which again compare well with experimentally observed blisters (in orange). In this model, the channel heights rather scales with $h_B \propto \Delta V^{1/3}$, which is also consistent with experimental data (see Fig. S24a).

Independent of the mechanism, our models predict a strong sensitivity of the phenomenon to surface charge. This prediction is corroborated experimentally by varying the salt concentration, which modulates the surface charge[30,41], and reproduces the expected trend (Fig. S22). This framework rationalizes the unidirectional positive blisters that underlie the memory of highly asymmetric channels[21]. While this previous work reported the formation of micrometre-scale blisters, we observe here much smaller and more diverse deformations in 2D nanochannels. We observed a majority of unidirectional blisters forming at positive voltages as in Figure 4f, but also found occurrences of negative unidirectional blisters such as the one presented in Figure 5a and Fig. S11a. We attribute this variability to the complexity of the layered structure and the possible local variations in the surface charge landscape. This points to a rich and device-dependent phenomenology that has probably been overlooked on other nanofluidic platforms. These results demonstrate that to fully capture the physics of nanofluidic transport, one must consider not only the forces acting on the confined molecules but also those exerted on the confining walls.

## Conclusion

In summary, we demonstrate that voltage-induced mechanical blistering drives current-voltage hysteresis and ionic memory in 2D nanochannels. While local ionic effects may occur, our analysis establishes physical deformation as the only scenario and primary state variable compatible with the observed memory (Supplementary Discussion). The stochastic and evolving nature of these deformations explains device-to-device variability and device evolution, and likely extends to other nanofluidic architectures. Our analysis identifies two electromechanical pathways: osmotic pressure from concentration polarization and electrostatic forces on surface charges, each capable of generating sufficient stress to overcome van der Waals adhesion. Our correlative approach rationalizes the behaviour of all the measured devices, highlighting the importance of operando imaging for understanding nonlinear molecular transport in nanofluidic systems. It shows that the assumption of static confining walls does not hold at high voltages, revealing an intrinsic coupling between mechanics and transport that parallels the conformational gating of biological ion channels[43]. By fully accounting for these behaviours, our approach provides the basis for overcoming variability challenges, a necessary step toward reliable nanofluidic technologies.

This success of operando imaging calls for its widespread use for understanding nonlinear ion transport at the nanoscale, in particular to reassess previous conclusions. Because a purely ionic memory mechanism would be expected to induce only minute refractive index variations (<1%), the resulting signal would likely fall below the detection limit of our technique, in stark contrast to its pronounced sensitivity to structural deformations. To address this limitation, complementary imaging approaches could be employed in operando in nanofluidic systems that do not exhibit mechanical distortions. For ultrasmall nanofluidic architectures, such as nanotubes or biological nanopores, single-molecule super-resolution microscopy represents a particularly promising strategy. Beyond these imaging frontiers, controlling and optimising nanofluidic blisters through mechanics and interfacial chemistry provides concrete design principles for ionic memory, for instance, through the deterministic placement of nanofabricated low-adhesion zones[44,45]. On this basis, high-performance nanofluidic synapses integrated into large-scale liquid circuits become a realistic prospect for advanced ionic computing.

## Acknowledgements

We acknowledge funding from the European Research Council (grants 101020445—2D-LIQUID N.R. and A.R.), the Swiss National Science Foundation

(grant TMPFP2-217134 to T.E.) and Swiss Government Excellence Scholarship to A.D.P.; B.R. acknowledges funding from the Royal Society University Research Fellowship renewal URF\R\231008, Philip Leverhulme Prize PLP-2021-262, European Union's H2020 Framework Programme\ERC Starting Grant 852674 – AngstroCAP, EPSRC new horizons grant EP/X019225/1. B.R. and A.K. acknowledge EPSRC strategic equipment grant EP/W006502/1. A.K. acknowledges Royal Society International Exchanges grant IES\R3\243222. B.C. acknowledges support from the CFM Foundation and the NOMIS Foundation. N.R. acknowledges support from an EMBO fellowship (ALTF1237-2024).

## Author Contributions

B.R., A.R., T.E. and N.R. designed and directed the project. S.K.V. fabricated the nanochannel devices with inputs from B.R. and A.K. S.K.V. performed the operando measurements and their analysis, with input from N.R. and T.E.. A.D.P. performed additional operando measurements. N.R. built the operando imaging microscope, and T.E. designed the fluidic cell. N.R. developed the Python script for image analysis, kymograph and statistical analysis. N.R. carried out TMM calculations. B.C. developed the theoretical modelling with input from N.R.; S.K.V., N.R., T.E. and B.C. wrote the paper with inputs from A.R., A.K. and B.R. All authors contributed to the discussion.

## Competing interests

The authors declare no competing interests.

# *Supplementary Information*

## Direct imaging elucidates ionic memory in two-dimensional nanochannels

Kalluvadi Veetil Saurav[1,2]*, Nathan Ronceray[3,9]*, Baptiste Coquinot[4], Agustin D. Pizarro[3], Ashok Keerthi[1,2,6], Theo Emmerich[3,5,#], Aleksandra Radenovic[3,7,#], Boya Radha[2,6,8,#]

*These authors contributed equally

#To whom correspondence should be addressed;
e-mail: theo.emmerich@ens-lyon.fr,
aleksandra.radenovic@epfl.ch, radha.boya@manchester.ac.uk

[1] Department of Chemistry, School of Natural Sciences, The University of Manchester, Manchester, M13 9PL, U.K.

[2] National Graphene Institute, The University of Manchester, Manchester, M13 9PL, U.K.

[3] Institute of Bioengineering, School of Engineering, EPFL, Lausanne, Switzerland

[4] Institute of Science and Technology Austria, Austria

[5] Laboratoire de Physique, UMR CNRS 5672, ENS de Lyon, Université de Lyon, Lyon, France

[6] Photon Science Institute, The University of Manchester, Manchester, M13 9PL, U.K.

[7] NCCR Bio-Inspired Materials, EPFL, Lausanne, Switzerland

[8] Department of Physics and Astronomy, School of Natural Sciences, The University of Manchester, Manchester, M13 9PL, U.K.

[9] Present address: Laboratoire Photonique Numérique et Nanosciences, Univ. Bordeaux, Talence, F-33400 France

# 1. Fabrication of 2D nanochannels

The fabrication of angstrom-channel devices (Fig S1a) involves stacking individual layers of 2D materials, such as graphene, hexagonal boron nitride into a stack on a 500nm thick double-sided polished silicon nitride on silicon ($SiN_x$) substrate. We followed a modified version of the fabrication process previously reported[1,2] and is outlined below (Fig S1b):

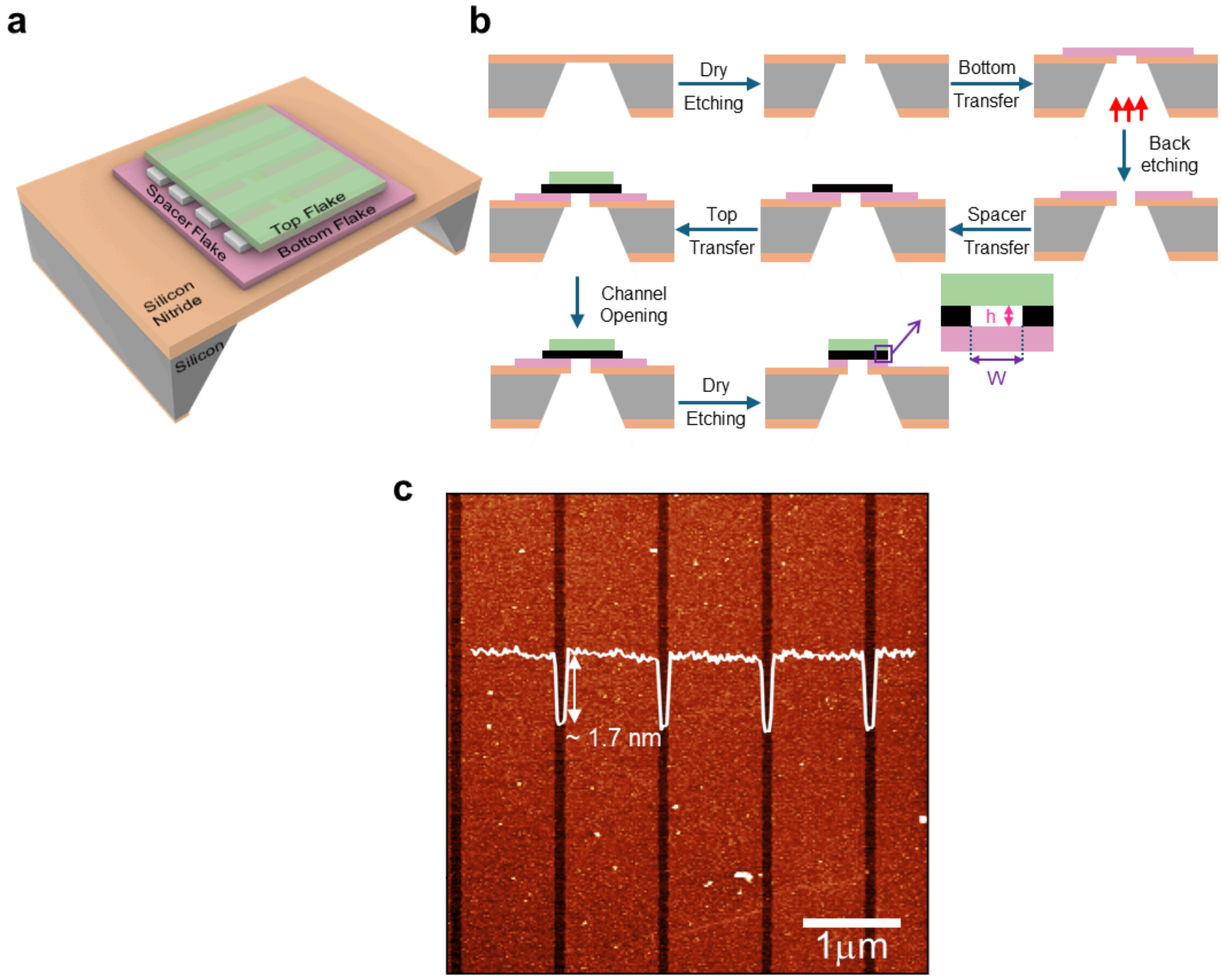

**Figure S1** (a) Schematic of a nanochannel device (b) Schematic of the fabrication step involved. A micro hole is made on the free-standing $SiN_x$ membrane by photolithography, followed by dry etching. Onto this substrate, a hBN bottom flake is transferred and back etched (dry etching). Subsequently, a top flake is transferred above the spacer. A spacer flake is made by the e-beam lithography technique, wherein a parallel array of strips is patterned, followed by oxygen plasma etching. Then the top-spacer stack is further transferred onto the bottom substrate. To define channel length, a rectangular aperture is made using photolithography, followed by dry etching. (c) AFM profile of a five-layer graphene spacer flake with a channel width of ~100 nm separated by ~1 μm.

The nanochannel device in this study features a trilayer stack, where spacers are sandwiched between bottom and top flakes made of 2D materials. The spacers, arranged as an array of parallel strips, serve as channels for ion transport. We chose hBN as the top material for its transparency, which allows light to pass through, providing insights into the dynamics within the trilayer stack. These stacks are assembled onto a silicon nitride substrate, and the channel lengths are defined by creating a rectangular aperture through photolithography, followed by reactive ion etching.

## 2. Imaging setup

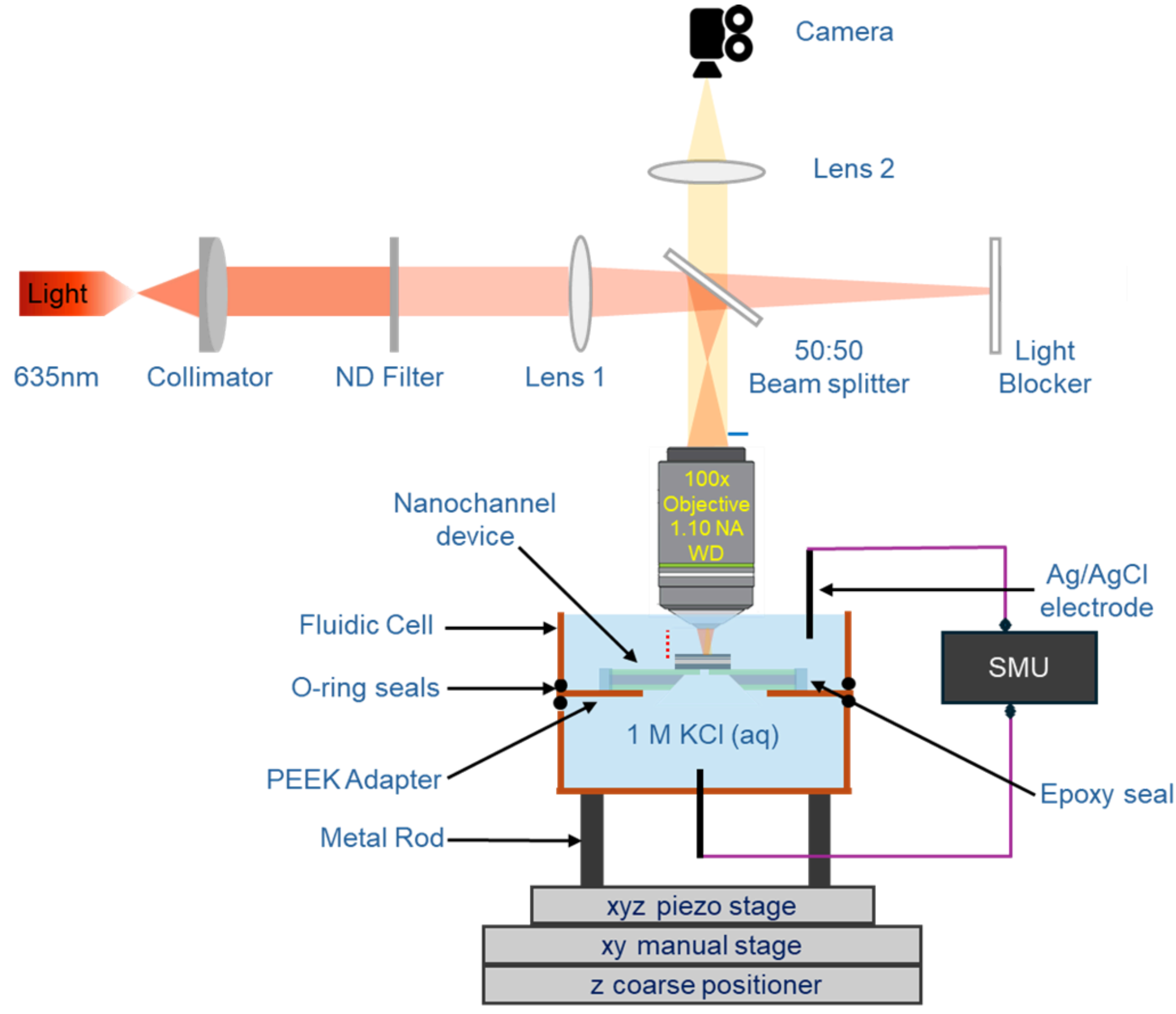

**Figure S2.** Schematic of the optofluidic setup. The setup consists of a white light source with a 10 nm bandpass filter with a central wavelength of 635 nm, focused on the back focal plane of a water-dipping objective, and further to the device immersed in an aqueous solution. The reflectance image is then acquired using a CMOS camera.

In-operando thin film interferometry measurements were performed using the experimental setup illustrated in Figure S2. Imaging was carried out using a water-dipping objective integrated into a custom-designed fluidic cell. Given that the dimensions of the nanofluidic chips were smaller than the water-dipping objective, a PEEK adapter (square with a 1 mm aperture) was fabricated to support the chip, which was affixed using epoxy resin (Araldite Standard). The chip was mounted in an upward orientation to facilitate imaging the trilayer stack on the $SiN_x$ substrate.

This adapter was positioned between the two components of the PEEK fluidic cell, which were secured by threaded connections and sealed using O-rings. The bottom chamber of the cell included a leak-proof opening to accommodate the bottom electrode, whereas the top chamber remained open to allow immersion of both the objective and the top electrode.

A Nikon CFI Plan 100X ($NA$=1.1) water-dipping objective with a 2.5 mm working distance was employed for imaging. The objective-to-sample distance was initially adjusted using a coarse z-positioner, followed by manual sample positioning via an xy translation stage. Fine adjustments for optimal focus were achieved using a 3-axis piezoelectric stage.

The sample was illuminated using a thermal white light source with a 10 nm bandpass filter having a central wavelength of 635 nm, and was attenuated using neutral density (ND) filters to achieve appropriate illumination power densities. The collimated beam was then focused onto the back focal plane of the objective using an achromatic doublet lens.

The objective was mounted at the base of a filter cube mount containing a 50/50 non-polarizing beam splitter. This configuration allowed 50% of the illumination light to be directed toward the sample, while the remaining 50% of the reflected light was transmitted towards the camera. Image acquisition was performed using a Hamamatsu ORCA Flash 4.0 CMOS camera, with an achromatic doublet lens employed for image formation.

# 3. Electrokinetic Measurements

Electrokinetic measurements were performed using Ag/AgCl electrodes to apply the potential and record the resulting current. The setup included a FEMTO amplifier (DLPCA-200) and a digital/analog converter (NI 63 series), which also synchronises the electrokinetic data with optical imaging. Conductance traces presented in Figures 3,4 and 5 were obtained by dividing current by voltage, omitting points near zero voltage with ill-defined conductance. The conductances $G_{min}$ and $G_{max}$ used to define the normalized conductance $\delta G$ were computed as the 0.1 and 0.9 quantiles of the conductance trace to average out the cycle-wise variability of the conductance minima and maxima.

# 4. Image acquisition and processing

In the case of drift during the acquisition, the image was stabilized using the image stabilizer plugin for ImageJ by Li and Kang[3]. The image was initially acquired with a 2×2 binning, followed by an additional 2×2 binning applied during data analysis.

# 5. Transfer matrix calculation of the contrast

## 5.1 Simulation details

The 6-layer thin film reflectance is evaluated using the Transfer Matrix Method (TMM). We consider normal-incidence reflection from a multilayer structure immersed in water, with the following stack (from top to bottom):

1. Semi-infinite water: $n_0=n_w$
2. Top hBN: thickness $t_{top}$, refractive index $n_1=n_{hBN}$
3. Water blister: thickness $h_B$, refractive index $n_2=n_w$
4. Bottom hBN: thickness $t_{bot}$, refractive index $n_3=n_{hBN}$
5. Silicon nitride (SiN): thickness $t_{SiN}$, refractive index $n_4=n_{SiN}$

6. Semi-infinite water: $n_5=n_w$

Our goal is to compute the reflected intensity $\eta(h_B(\Delta V))$ to obtain the VIC. We assume no losses in the media and that the resulting voltage-induced intensity modulations arise purely from interference. Moreover, we model the case of blisters much wider than the light wavelength, neglecting refraction effects at the edges.

Interface and propagation matrices

Interface and propagation through the layers of the heterostructure can be formalized through transfer matrices acting on the electric field vector, $\begin{pmatrix} E^{\downarrow} \\ E^{\uparrow} \end{pmatrix}$ where $E^{\downarrow}$ corresponds to the downward-propagating wave and $E^{\uparrow}$ to the upward-propagating wave.

At each interface between refractive indices $n_i$ and $n_j$, we define the reflection coefficients $r_{ij}=(n_i-n_j)/(n_i+n_j)$ and transmission coefficients: $t_{ij}=2n_i/(n_i+n_j)$.

An interface from medium i to j contributes an interface matrix $I_{ij}$, and a layer of thickness $t$ and index $n$ contributes a propagation matrix $P(n, t, \lambda)$ , respectively given by:

$$I_{ij} = \frac{1}{t_{ij}} \begin{pmatrix} 1 & r_{ij} \\ r_{ij} & 1 \end{pmatrix} \quad ; \quad P(n, t, \lambda) = \begin{pmatrix} exp(2i\pi nt/\lambda) & 0 \\ 0 & exp(-2i\pi nt/\lambda) \end{pmatrix}$$

We now construct the full transfer matrix $M(h_B)$ for the 6-layer system. The light is incident from the top (semi-infinite water), and propagates through each interface and layer in sequence.

$$M(h_B)=\begin{pmatrix} M_{11} & M_{12} \\ M_{21} & M_{22} \end{pmatrix} = I_{01} \cdot P(n_1,t_{top}) \cdot I_{12} \cdot \boldsymbol{P(n_2,h_B,\lambda)} \cdot I_{23} \cdot P(n_3,t_{bot}) \cdot I_{34} \cdot P(n_4,t_{SiN}) \cdot I_{45}$$

In this expression, we highlighted $\boldsymbol{P(n_2,h_B,\lambda)}$ which is the only blister-dependent part.

Reflected Intensity $\eta(h_B)$

In the absence of incoming light from below (boundary condition $E^{\uparrow}=0$ in the bottom water), the reflection coefficient $r(h_B)$ is given by: $r(h_B)=M_{21}(h_B)/M_{11}(h_B)$

This yields the reflected intensity as: $\eta(h_B)=|r(h_B)|^2$ and thus the contrast through equation (1).

## 5.2 Simulation results

Results are shown in Figure S3, with the aim of demonstrating that semi-quantitative topographical information can be retrieved from contrast maps *in the device geometries used in this work*. The two main geometrical parameters that can be set experimentally are the bottom and top thicknesses $t_{bot}$ and $t_{top}$ chosen during the van der Waals heterostructure fabrication. We examined the parameter range in which the contrast scales monotonously with the blister height $h_B$.

Figure S3d presents the TMM-evaluated contrast for a range of blister heights from 1 nm to 80 nm. We find that larger blisters provide an increasing contrast, which can be either negative (blue) or positive (red), depending on the region of the parameter space $\{t_{bot}, t_{top}\}$ chosen during fabrication. Some parameter choices result in reflectance cancellation, as shown in S3b. These regions should be avoided, as the contrast then provides ambiguous results over the device topography. For this reason, the light wavelength and device geometry should be chosen to avoid this pathological case. Thus, we used λ=635 nm and we operated in the geometry range given by the yellow square in Figure S3d ($t_{bot}$=50±5 nm and $t_{top}$=90±10 nm), for which the blister formation always results in a positive contrast that scales monotonously with the blister height (Fig. S3b). The contrast-to-height relationship is sub-linear but can be approximated by a linear increase near zero, which defines the deformation sensitivity $F$, as defined in equation (2). The distribution of $F$ values in the used geometry range has a median value of 1.55 with extremal values of 1.40 and 2.27 (Fig. S3c). This shows that for small deformations, the VIC gives semi-quantitative information over the blister height within a factor ~2, and can be made fully quantitative by accurate AFM height measurements of the device.

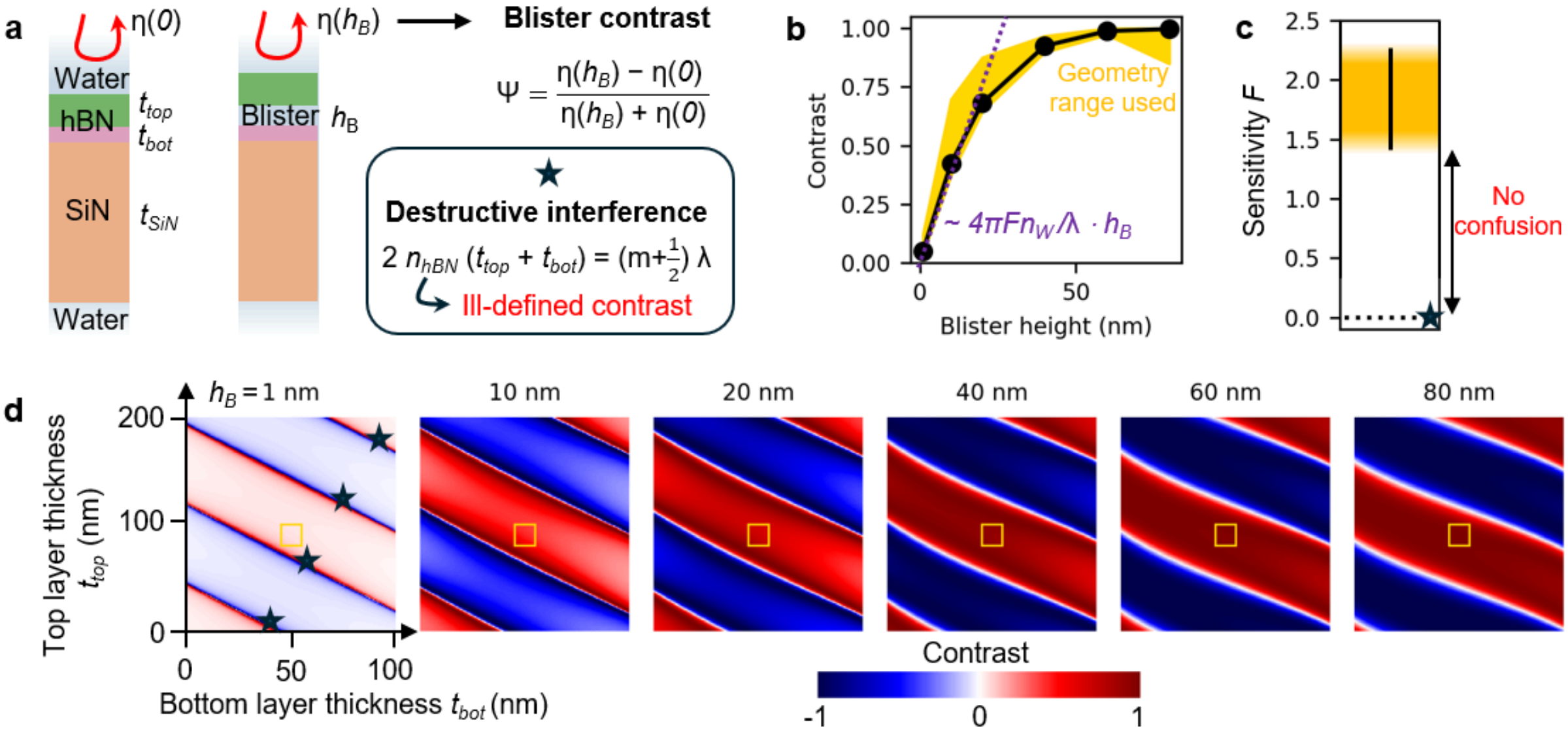

**Figure S3. Transfer matrix method (TMM) calculations of the optical contrast in the used device geometries. a**, Sketch of the thin film structure at rest and in the presence of a blister, used to compute the VIC Ψ. In the pathological case of destructive interference, the contrast is ill-defined, which necessitates carefully chosen experimental parameters. **b**, Contrast vs blister height curve showing the derivation of the sensitivity from the near-zero slope. **c**, Sensitivity values extracted from the used geometry range show that the sensitivity is always a positive number exceeding 1.4, thereby providing unambiguous information about small deformations. **d**, TMM contrast map for varying bottom and top layer thicknesses, given for blister heights ranging from 1 nm to 80 nm. The parameters used correspond to the yellow box in which the formation of a blister always results in a positive contrast change in red. The presence of singularities in the contrast along tilted lines corresponds to the destructive interference condition in **a**, leading to ill-defined contrast and sensitivity.

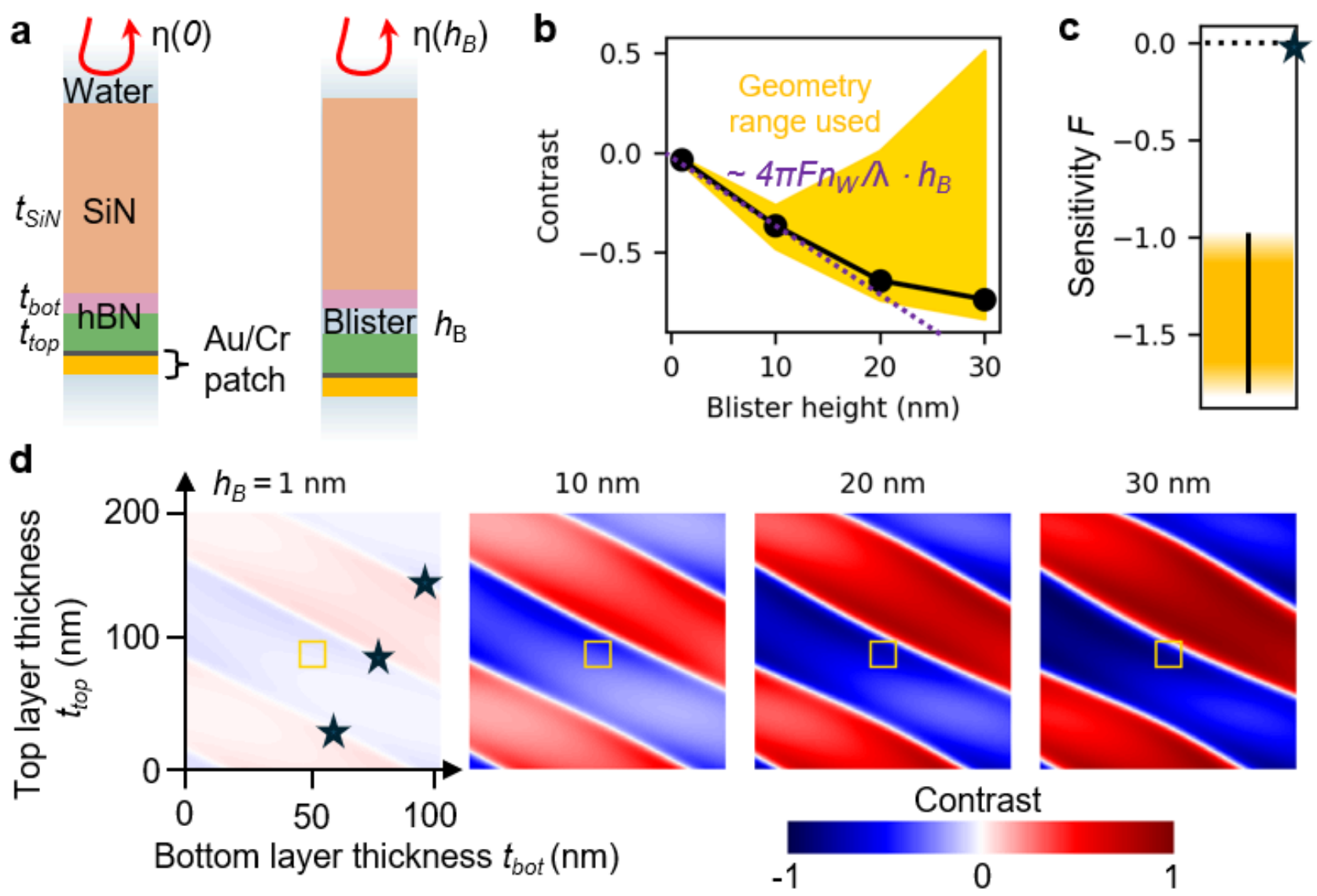

**Figure S4. Transfer matrix calculations of the optical contrast for gold patch devices imaged upside down. a**, Sketch of the thin film structure at rest and in the presence of a blister, for a device placed upside down and including a gold/chromium patch. **b**, Contrast vs blister height curve showing negative sensitivity. **c**, Sensitivity values extracted from the used geometry range show that the sensitivity is always a negative number below -1.0, thereby providing unambiguous information about small deformations. **d**, TMM contrast map for varying bottom and top layer thicknesses, given for blister heights ranging from 1 nm to 30 nm. The parameters used correspond to the yellow box in which the formation of a blister always results in a negative contrast change in blue, for small enough deformations.

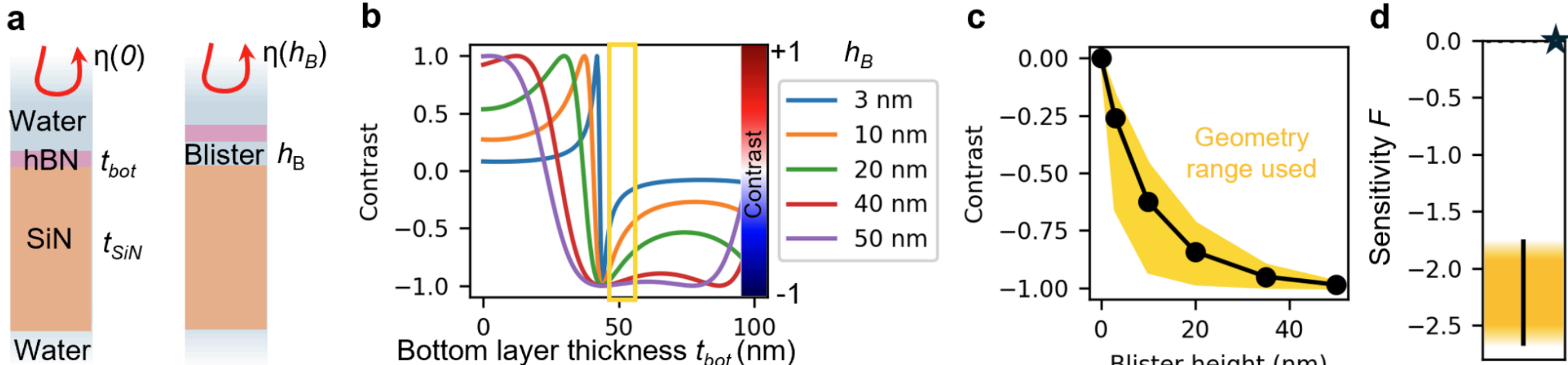

**Figure S5. Transfer matrix calculations of the optical contrast for topless parts of the devices. a**, Sketch of the thin film structure at rest and in the presence of a blister forming between the bottom hBN and the $SiN_x$ substrate. **b**, Contrast vs bottom layer thickness curve for different blister heights. **c**, Contrast vs height curve showing negative sensitivity for geometries used in this work. **d**, Sensitivity values extracted from the used geometry range show that the sensitivity is always a negative number below -1.8, thereby providing unambiguous information about small deformations.

# 6. Device information

## 6.1 List of results per device.

Table S1

| Device Number | Concentration | Species | Configuration | | | Height (nm) | Annealed (Yes/No) | Comments |
|---|---|---|---|---|---|---|---|---|
| | | | Top | Spacer | Bottom | | | |
| Device 1 | 1 M | KCl | hBN | Gr | hBN | 1.70 | No | |
| Device 2 | 1 mM - 1 M (KCl) | KCl, NaCl, LiCl, $CaCl_2$ | hBN | Gr | hBN | 0.68 | No | |
| Device 3 | 1 mM - 1 M (KCl) | KCl, NaCl, LiCl, $CaCl_2$, $AlCl_3$ | hBN | Gr | hBN | 1.02 | No | |
| Device 4 | 1 mM - 1 M | KCl, $CaCl_2$, $AlCl_3$ | hBN | Gr | hBN | 0.68 | Yes | |
| Device 5 | 1 M | KCl | hBN | Gr | hBN | 1.70 | Yes | Gold patch |
| Device 6 | 1 M | KCl | hBN | Gr | hBN | 1.02 | Yes | Gold patch |
| Device 7 | 1 M | KCl | hBN | hBN | hBN | 50 | Yes | Thick channel |
| Device 8 | 1 M | KCl | hBN | Gr | hBN | 1.02 | Yes | |
| Device 9 | 1 M | KCl | hBN | Gr | Gr | 1.7 | No | Bottom Gr |
| Device 10 | 1 M | KCl | hBN | Gr | Gr | 0.68 | No | Bottom Gr |
| Device 11 | 1 M | KCl | hBN | --- | hBN | 0 | Yes | |
| Device 12 | 1 M | KCl | --- | --- | hBN | --- | No | hBN on SiNx |
| Device 13 | 1 M | KCl | --- | --- | hBN | --- | No | hBN on SiNx |
| Device 14 | 1 M | KCl | hBN | Gr | Gr | 1.02 | No | Bottom Gr |

## 6.2 Device Endurance

All devices listed in Table S1 were measured multiple times, with each measurement having at least 2 cycles. We performed voltage amplitude sweeps on device 1 in the range of 5 mV to 1 V at 10 mHz and 100 mHz, followed by frequency variation studies. For device 2, we were able to complete the full experimental set, including amplitude, frequency, concentration, and salt-dependence studies and repeated the entire measurement three times to ensure reproducibility. All other devices we treated with at least 1 M KCl multiple times.

# 7. Nanofluidic measurements

## 7.1 Effect of ionic species

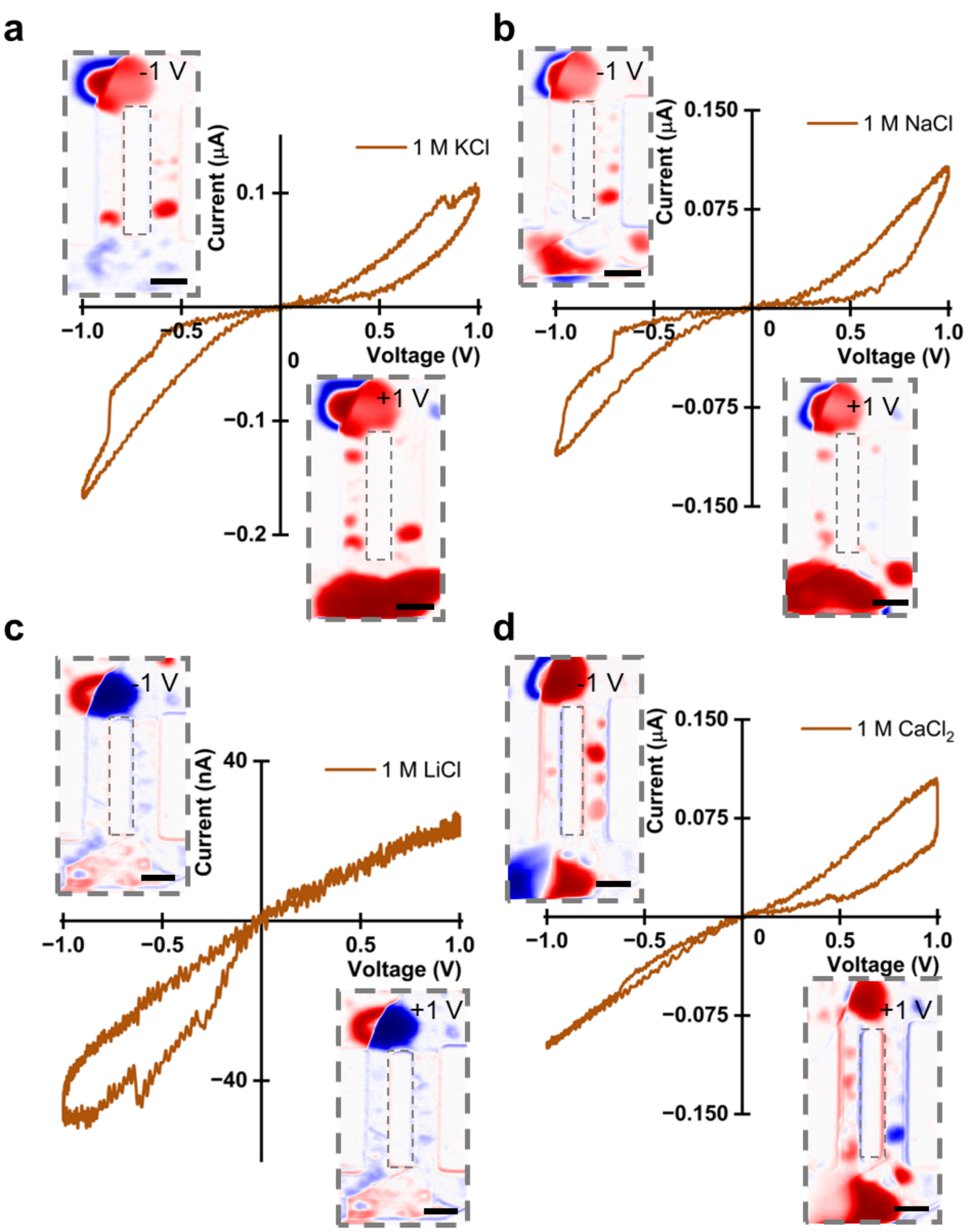

**Figure S6.** *I-V* curves and VIC images of device 2 at ±1 V with varying salt type **(a)** 1 M KCl, **(b)** 1 M NaCl, **(c)** 1 M LiCl and **(d)** 1 M $CaCl_2$ at a frequency $f$ = 10 mHz. The *I-V* response and VIC differ across salt types with KCl and NaCl showing bidirectional hysteresis (current loops in both voltage polarity) and LiCl and $CaCl_2$ showing uni-directional hysteresis (current loops in either of the voltage polarity). Scale bar: 5 µm

## 7.2 Voltage threshold for memristive effect

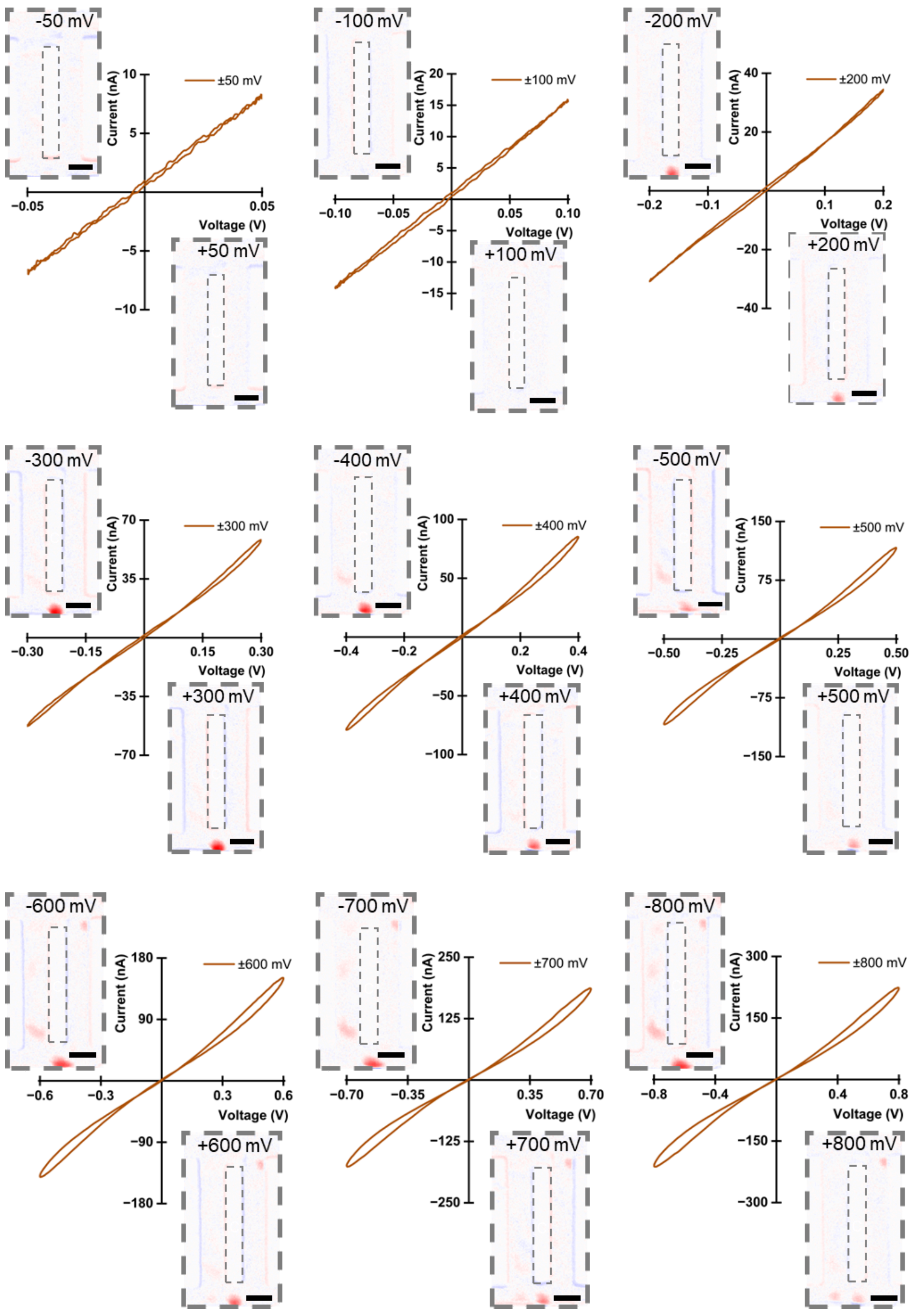

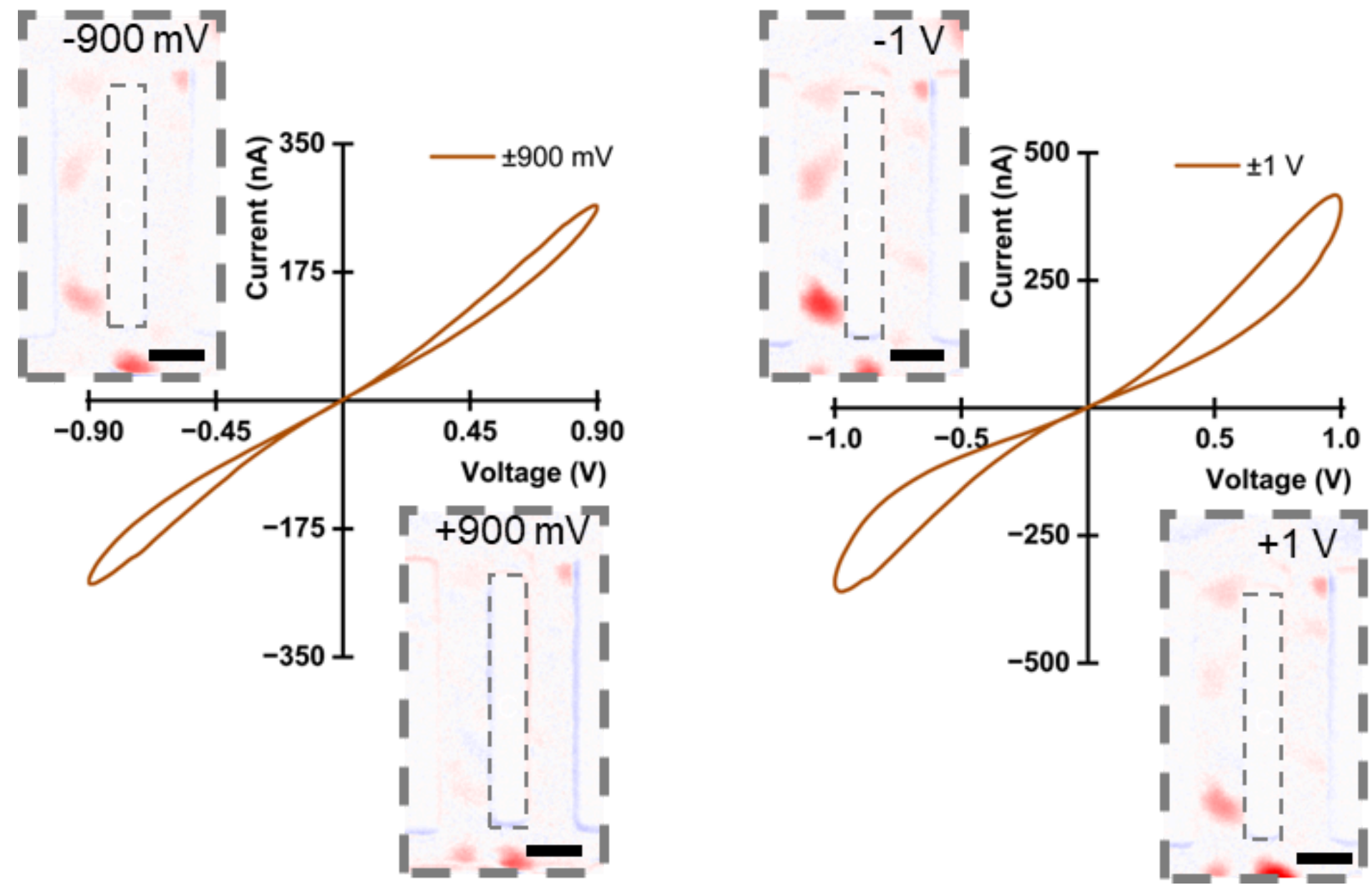

**Figure S7. Voltage Threshold.** *I-V* curves and VIC images of device 1 at voltage extremes with varying potential range (50 mV-1 V) in 1 M KCl at a frequency $f$ = 100mHz. Below threshold voltage ±0.2 V, *I-V* is linear and the VIC images show no contrast. As the voltage amplitude increases beyond ±0.2 V, the device shows hysteresis and VIC contrast appears and becomes more pronounced at higher voltage amplitudes. Scale bar: 5 µm

## 7.3 Additional measurements

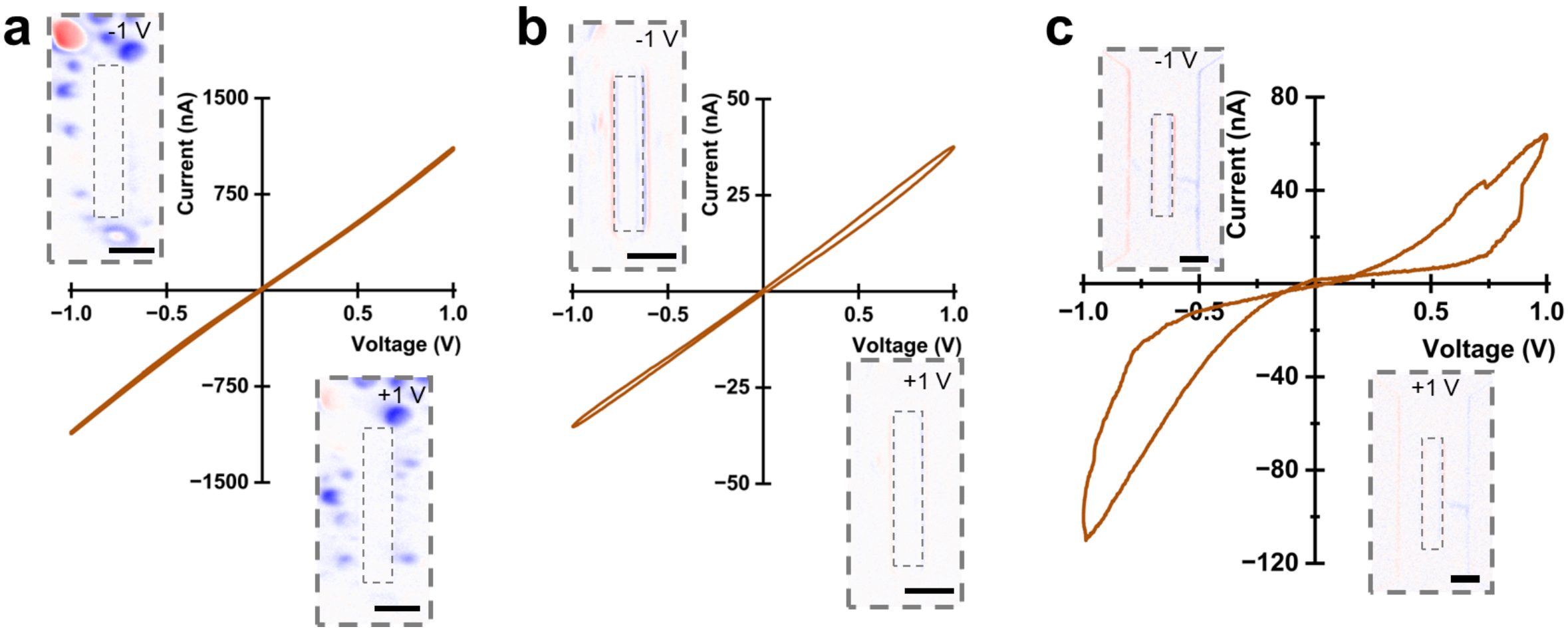

**Figure S8.** *I-V* curves and VIC images at -1V and +1V of (a) device 12 (only bottom hBN, with no spacer or top) in 1 M KCl at a frequency $f$ = 10 mHz. (b) device 13 (only bottom hBN, with no spacer or top) in 1 M KCl at a frequency $f$ = 100 mHz and (c) device 5 (gold patch device) at -1 V and +1 V in 1 M KCl at a frequency of $f$ = 10 mHz. (Here, the imaging is done from the bottom side of the device). Scale bar: 5 µm

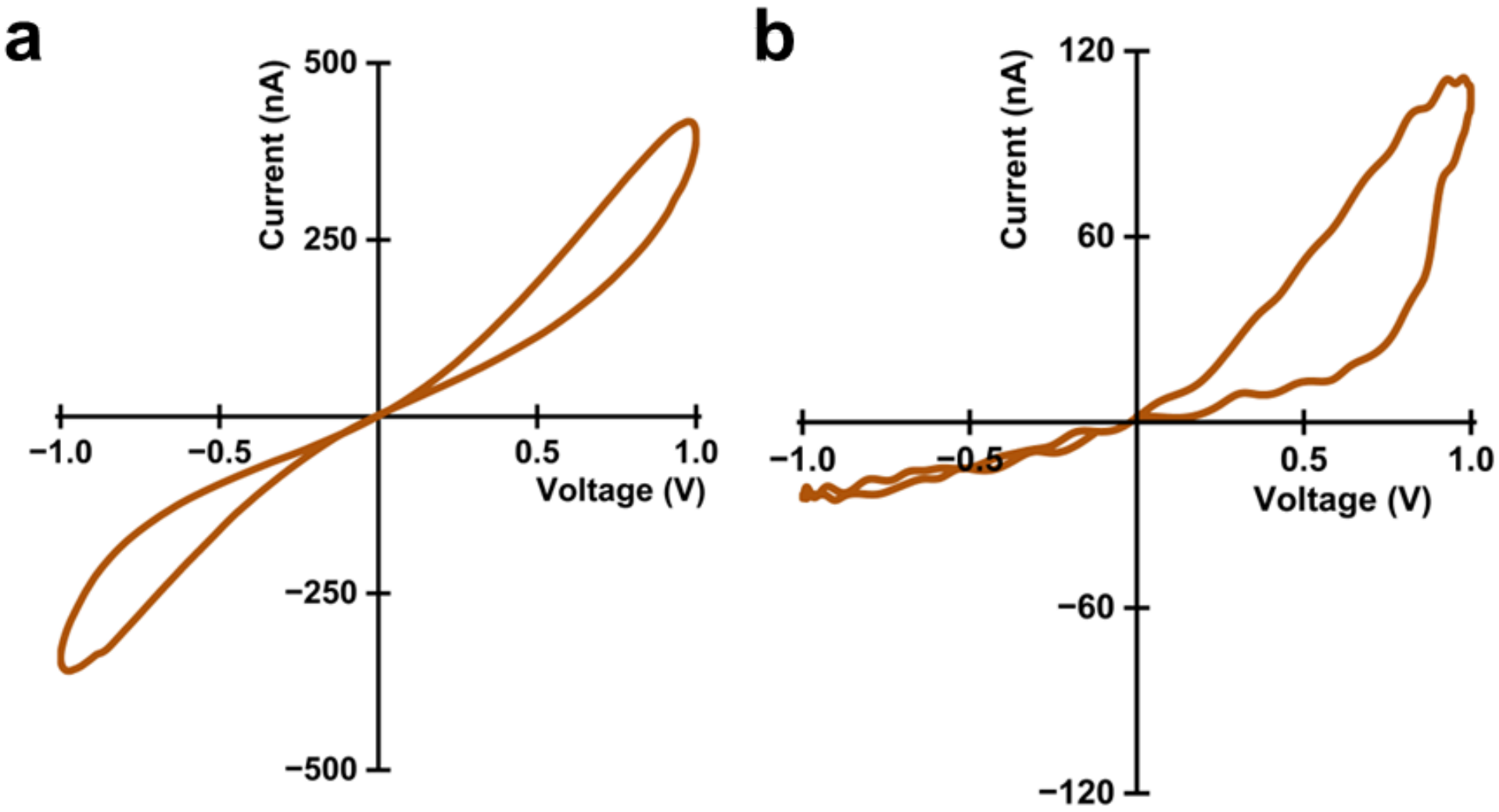

**Figure S9**. *I-V* curves corresponding to VIC images presented in figures **(a)** 4a from device 1 and **(b)** 4b from device 2. Device 1 shows bidirectional hysteresis with current loops in both voltage polarity. Device 2 shows unidirectional hysteresis with loop confined to only positive voltage polarity. For device 2, in the negative voltage polarity where no blister deformation is observed corresponds to a strictly linear off-state response similar to the low voltage observations in Fig. 2a reinforcing that hysteresis requires active structural deformation.

## 7.4 Conductance–contrast characteristics of bidirectional and unidirectional blisters

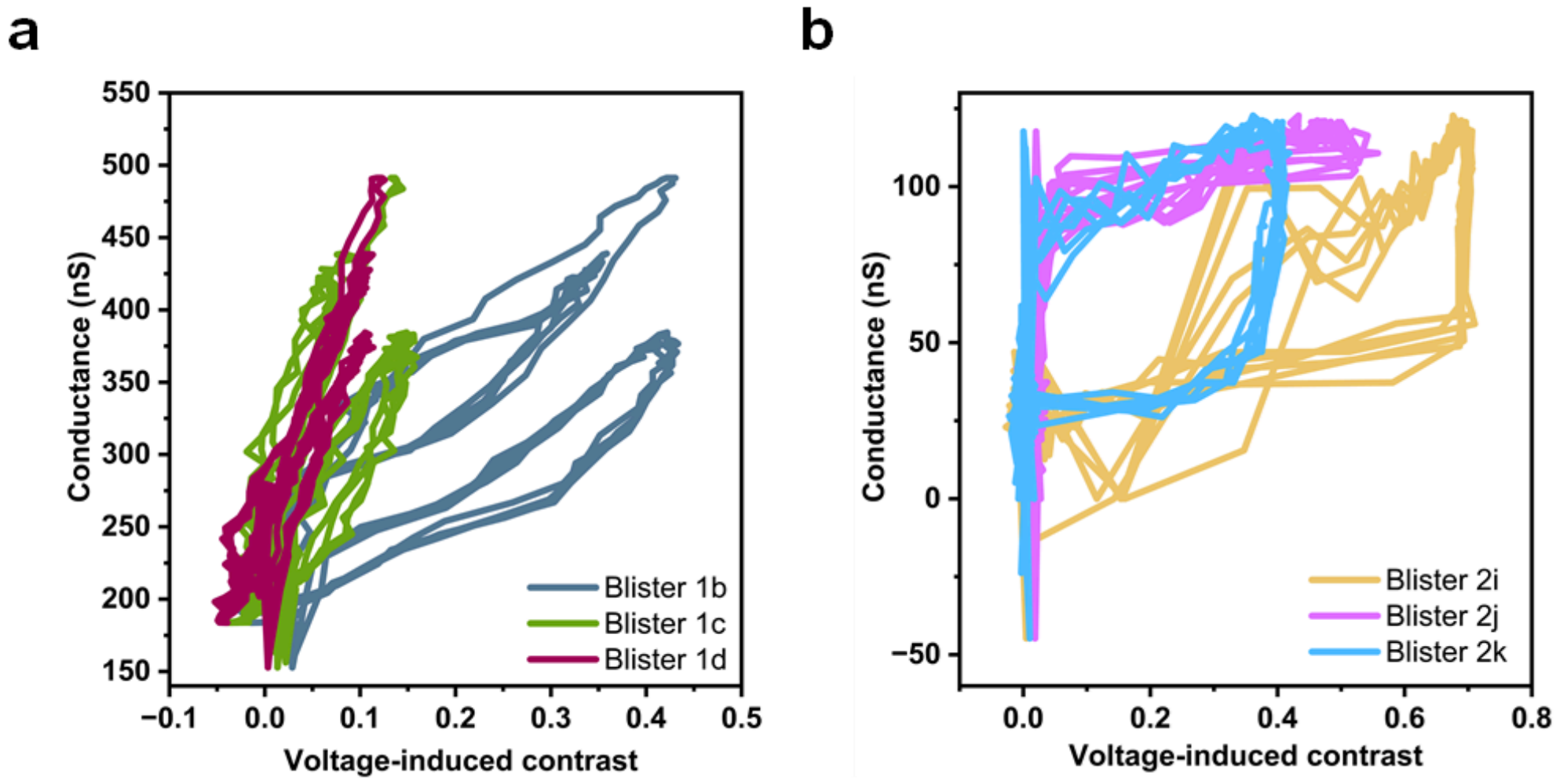

**Figure S10.** Conductance vs voltage induced graph for the (a) bidirectional hysteresis memristive device 1 shown in figure 4a and (b) unidirectional hysteresis device 2 shown in figure 4b.

## 7.5 Control devices measurements

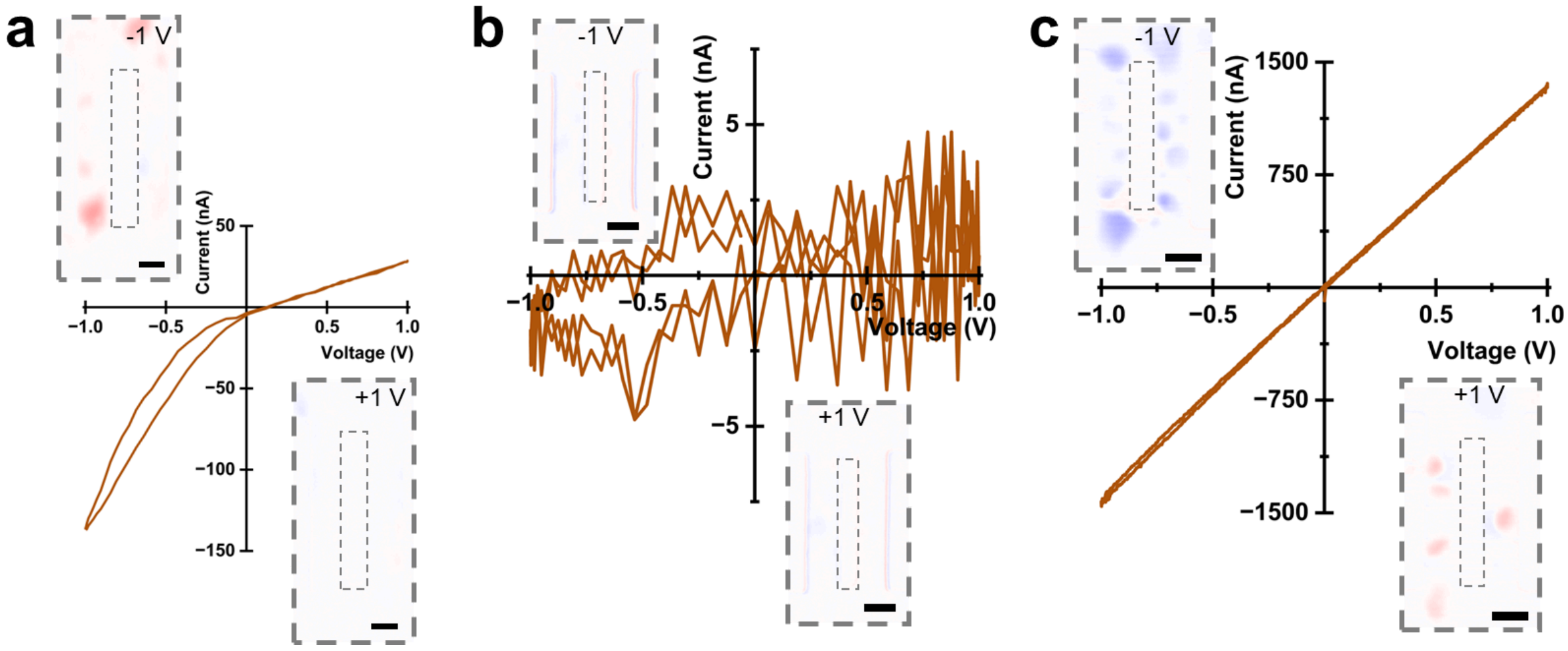

**Figure S11**. *I-V* curves and corresponding and VIC images at ±1 V of (a) device 4 at -1 V and +1 V, showing blister responsiveness to negative voltage in 1 M KCl at a frequency *f* = 100 mHz. The device shows VIC change predominantly at negative voltage range. (b) Device 11 (hBN as top and bottom and without any spacer) in 1 M KCl at a frequency *f* = 100 mHz with no VIC change at either voltage polarity and (c) device 7 (50 nm hBN spacer height ) in 1 M KCl at a frequency *f* = 10 mHz showing VIC change at both voltage polarity ranges with no hysteresis. Scale bar: 5 μm

## 7.6 Effect of frequency

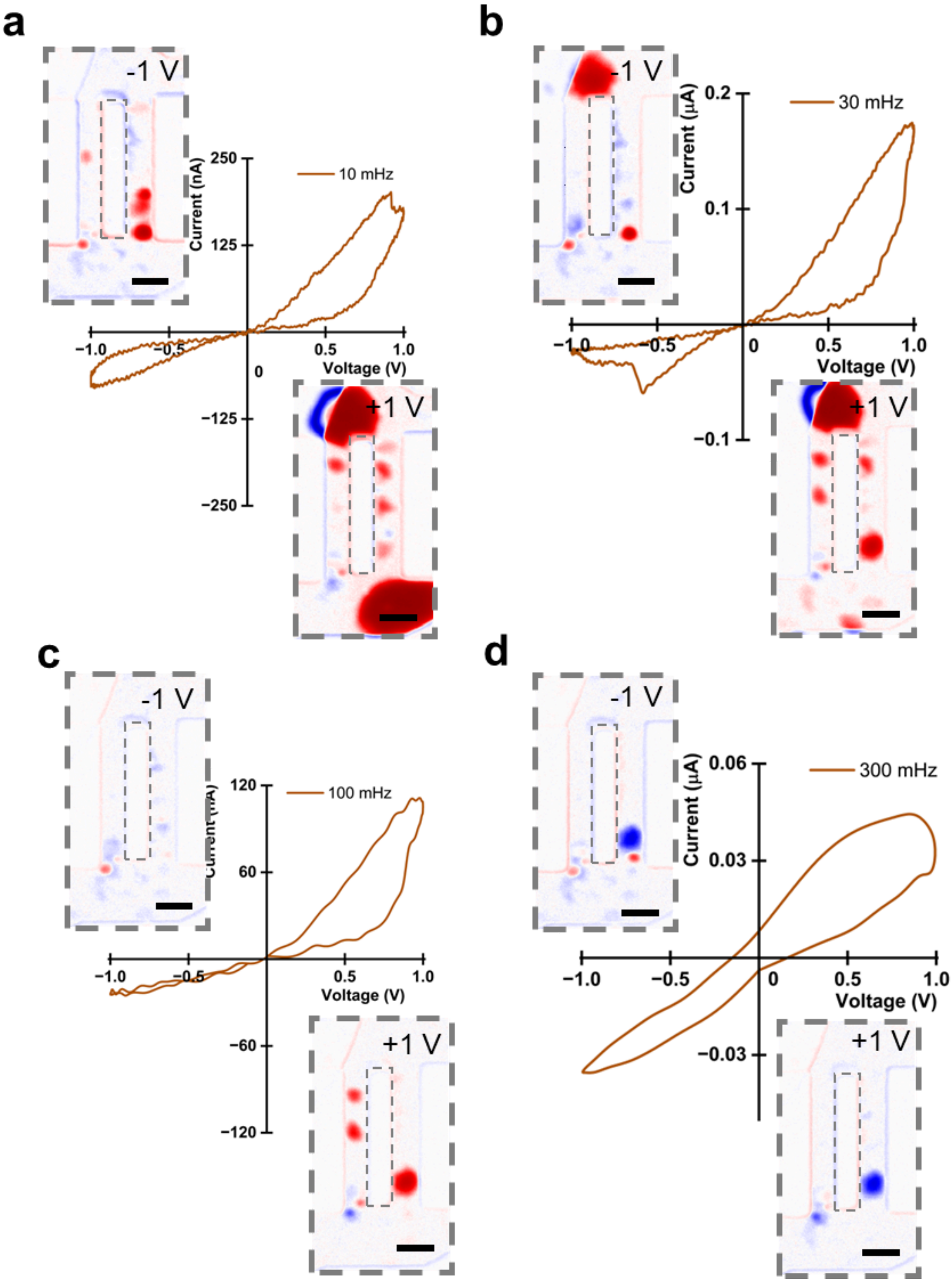

**Figure S12.** *I-V* curves and VIC images of device 2 at ±1 V with a varying frequency from 10 mHz to 300 mHz in 1 M KCl. The hysteresis transitions from bidirectional hysteresis to unidirectional at intermediate frequency and then becomes completely capacitive at 300 mHz with VIC diminishing across panels . Scale bar: 5 µm

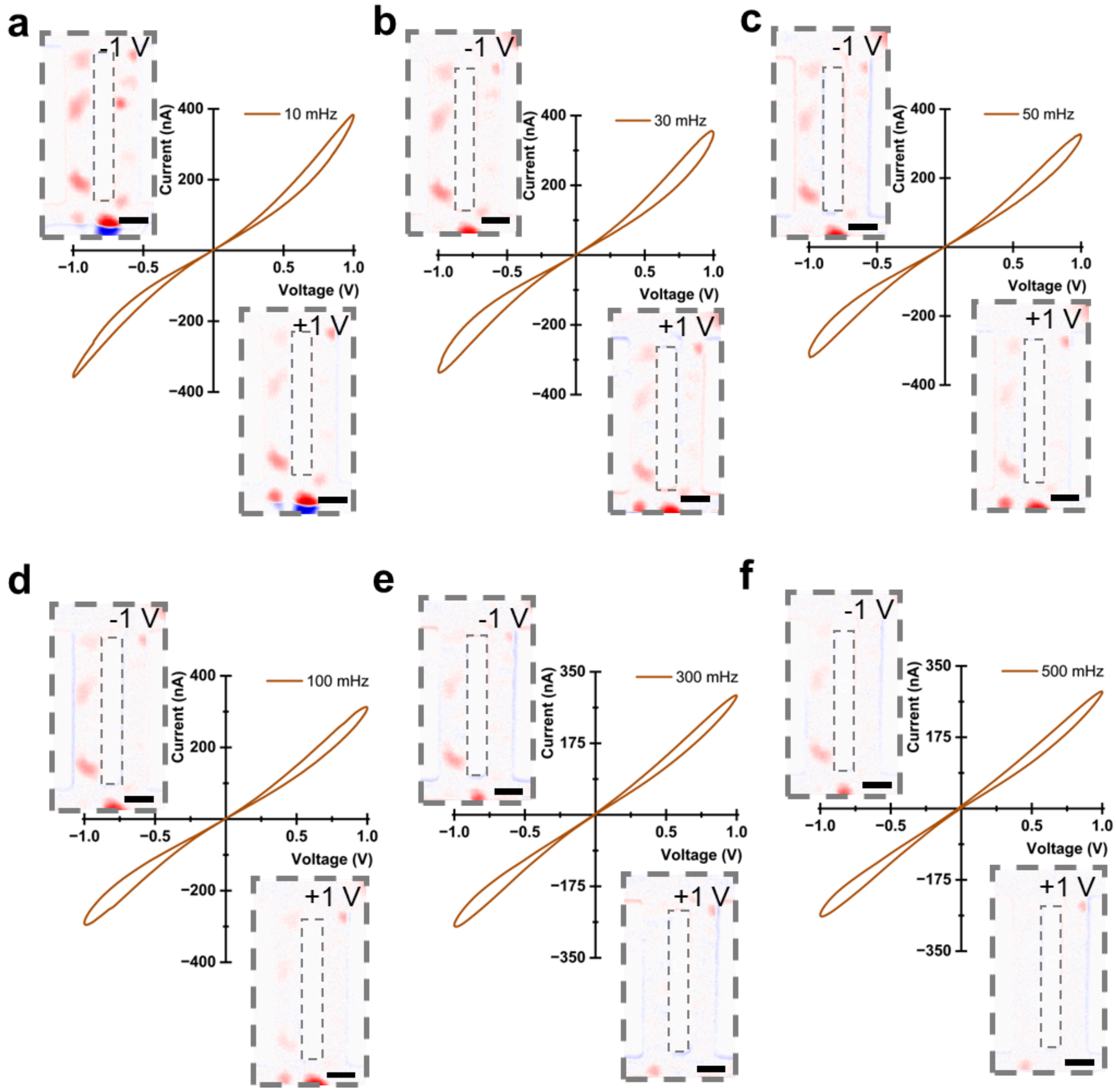

**Figure S13.** *I-V* curves and VIC images of device 1 at ±1 V with varying frequency (10 mHz to 500 mHz) in 1 M KCl. The bidirectional hysteresis loop reaches maximum at 30 mHz and narrows progressively with increasing frequency and the VIC diminishes correspondingly. The device remains bidirectional across the full frequency range. Scale bar: 5 μm

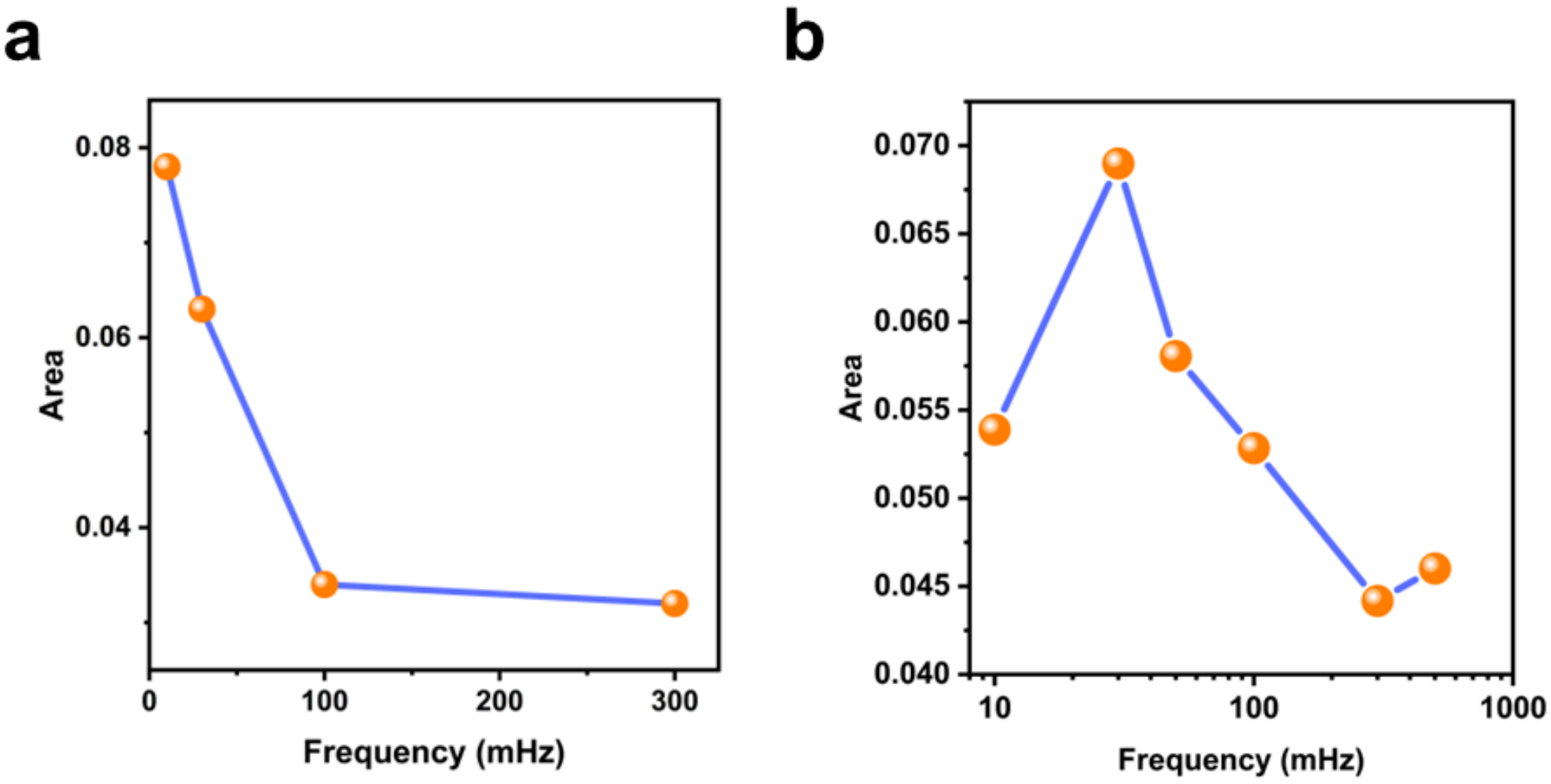

**Figure S14.** Hysteresis loop area as a function of frequency for (a) Device 2 and (b) Device 1, measured in 1 M KCl. For device 2, the loop area decreases monotonically. Device 1 shows a peak loop area 30 mHz before declining at higher frequencies. The corresponding *I-V* curves and VIC images for device 2 and device 1 are shown in Figure S12 and S13, respectively. In both the devices, the highest loop area is observed at lower frequencies.

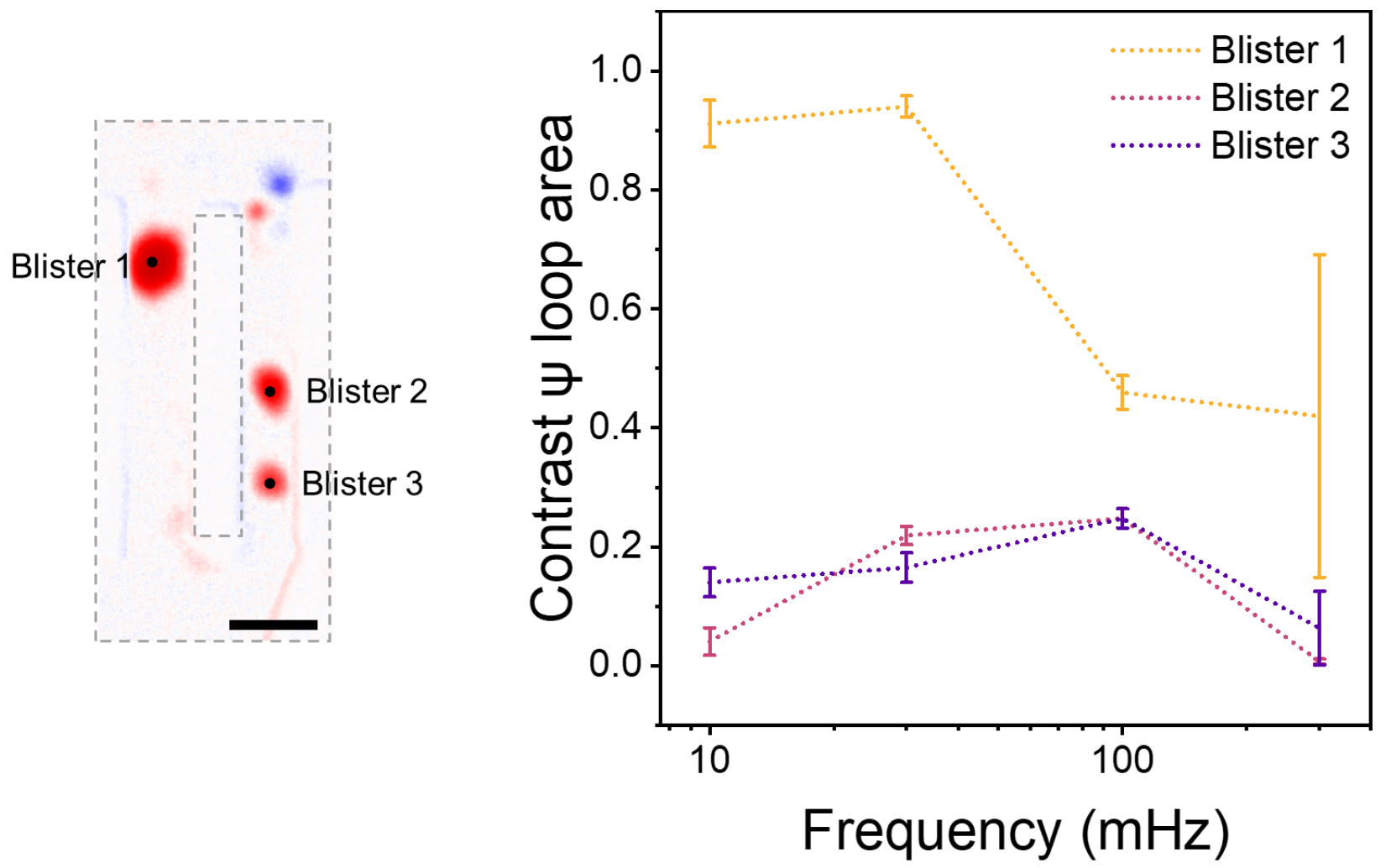

**Figure S15.** Per-blister breakdown VIC loop area vs frequency for selected blisters of device 2, measured in 1 M KCl with error bar from multiple area cycles (For *f = 10 mHz*; 2 cycles, *f = 30 mHz*; 2 cycles, *f = 100 mHz*; 6 cycles, *f = 300 mHz*; 6 cycles). The left panel shows a VIC image of device 2 with three blisters labelled. For blister 1, the maximum contrast loop area is observed at 30 mHz whereas for blister 2 and 3, it occurs at 100 mHz. Scale bar: 5 μm

## 7.7 Retention measurements

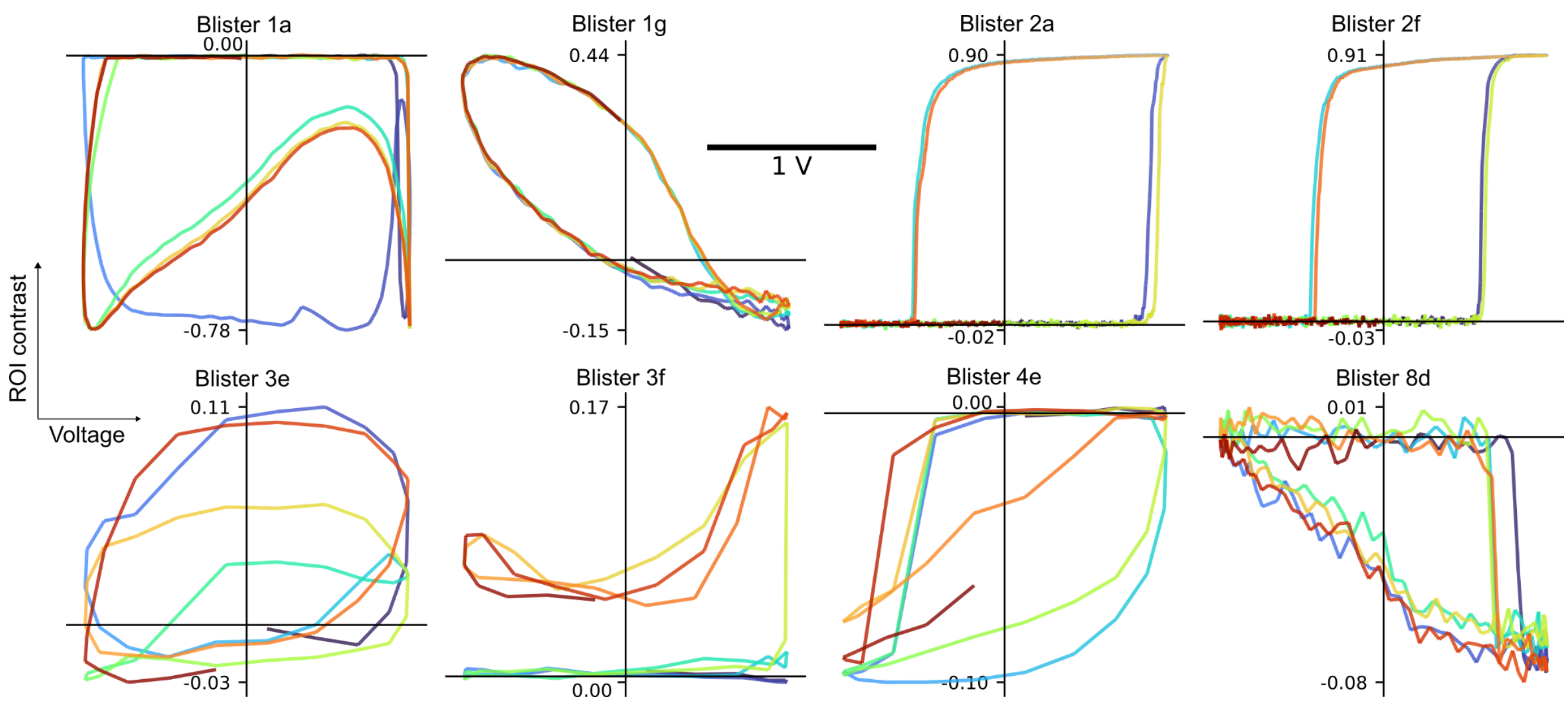

**Figure S16. Contrast-voltage curves of blisters exhibiting contrast retention at 0V, 8 occurrences out of 47 measurements.** While the majority of blisters fully relax to their baseline state at 0 V (N=39, non-retentive), a specific subset (N=8, presented here) maintains a significant structural deformation upon voltage removal, providing the physical basis for non-volatile memory behaviour. We note that all such retentive blisters identified in our analysis are unidirectional.

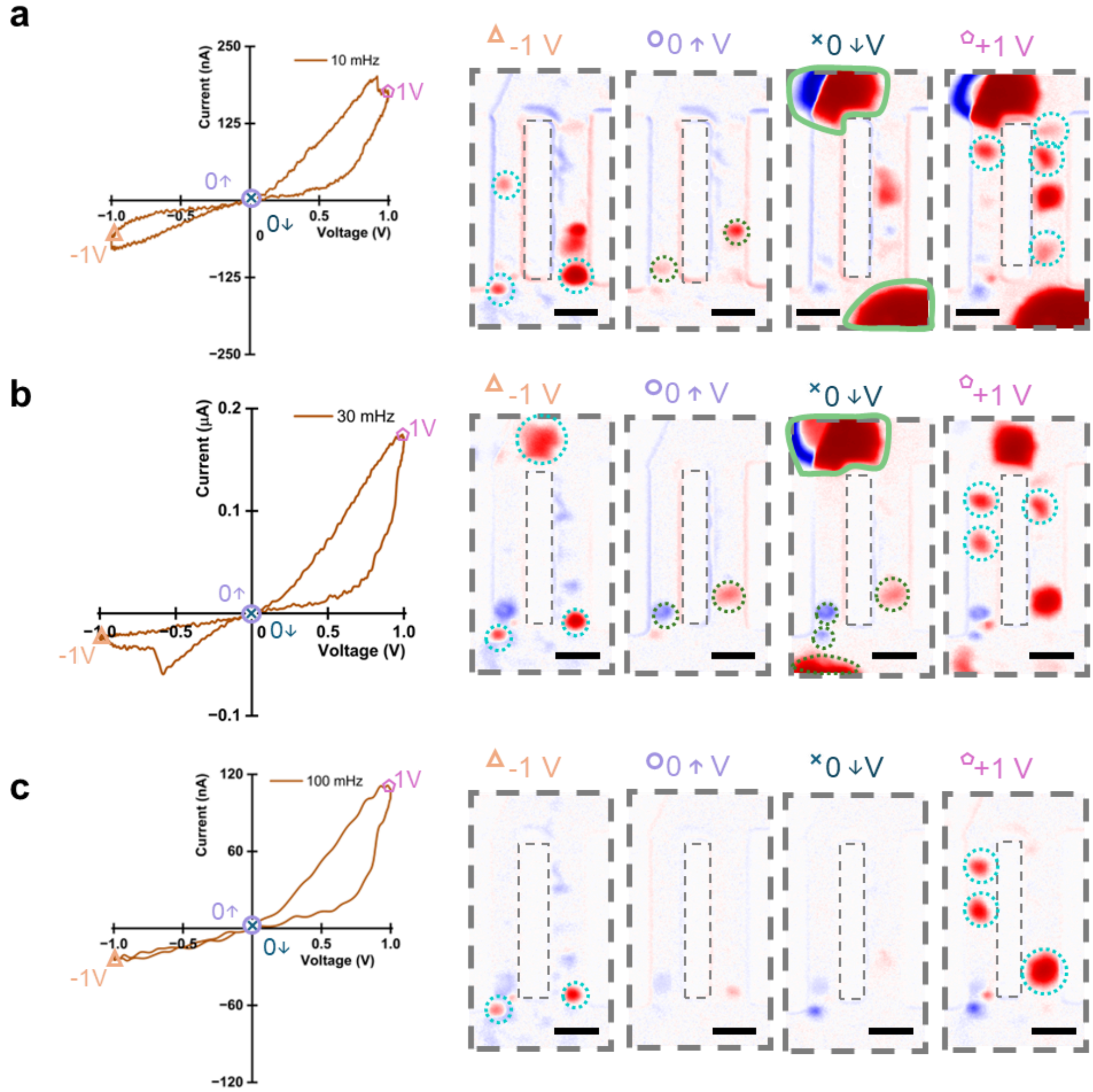

**Figure S17. Frequency-dependent transition between volatile and non-volatile structural states. a–c,** Operando VIC images and synchronized *I–V* curves for device 2 at driving frequencies of (**a**) 10 mHz, (**b**) 30 mHz, and (**c**) 100 mHz. Images are captured at voltage extrema (±1 V) and zero-crossings (0↓ V and 0↑ V). At lower frequencies (10-30 mHz), large blisters (light green outlines) persist after bias removal, resulting in non-volatile memory. In contrast, smaller blisters (cyan outlines) collapse immediately upon bias removal. At 100 mHz (**c**), the system transitions to a purely volatile regime where no retentive blisters are observed, directly corresponding to the suppression of the hysteretic state at the 0 V crossing in the electrical transport data. Scale bar: 5 µm.

## 7.8 Device evolution studies

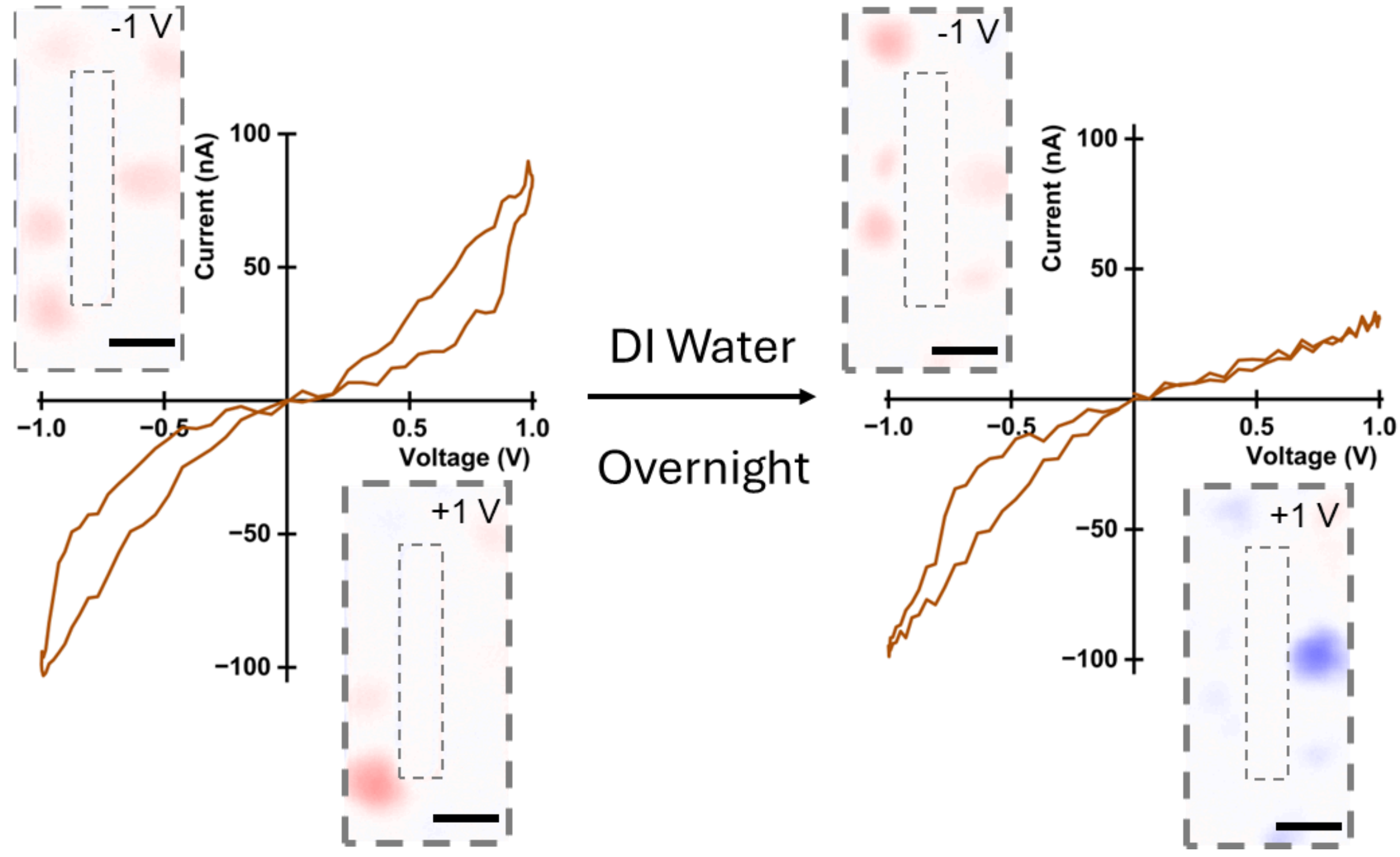

**Figure S18.** *I-V* curves and VIC images of device 3 at ±1 V in 1 M KCl at a frequency *f* = 100 mHz. The left panel shows the device before overnight submersion in DI water, where the *I-V* response exhibits bidirectional hysteresis. After leaving the device submerged in water overnight, the hysteresis shifts to unidirectional as shown in the right panel and VIC images at both the bias polarities change accordingly . Scale bar: 5 µm

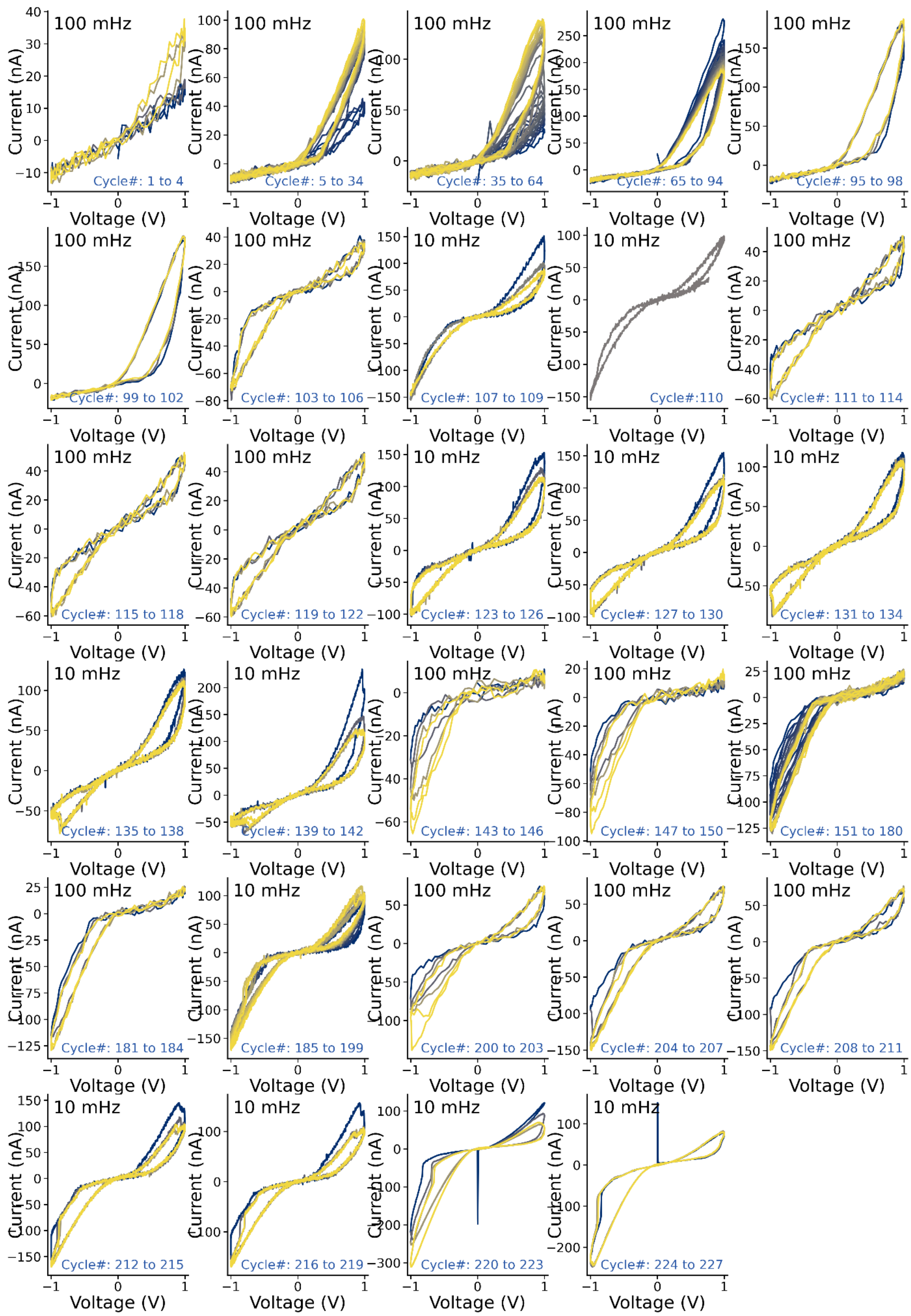

**Figure S19.** *I-V* cycles of device 2 over 227 measurement cycles in 1 M KCl, recorded at ±1 V and frequencies of 10 mHz and 100 mHz. Each panel shows a consecutive group of cycles, with earlier cycles in grey and later cycles in yellow. The dataset captures the full electrochemical history of the device providing a comprehensive picture of how memristive behaviour in nanochannel evolves under prolonged, repeated electrical stimulation. *I-V* traces are coloured blue to yellow in order of increasing cycle number.

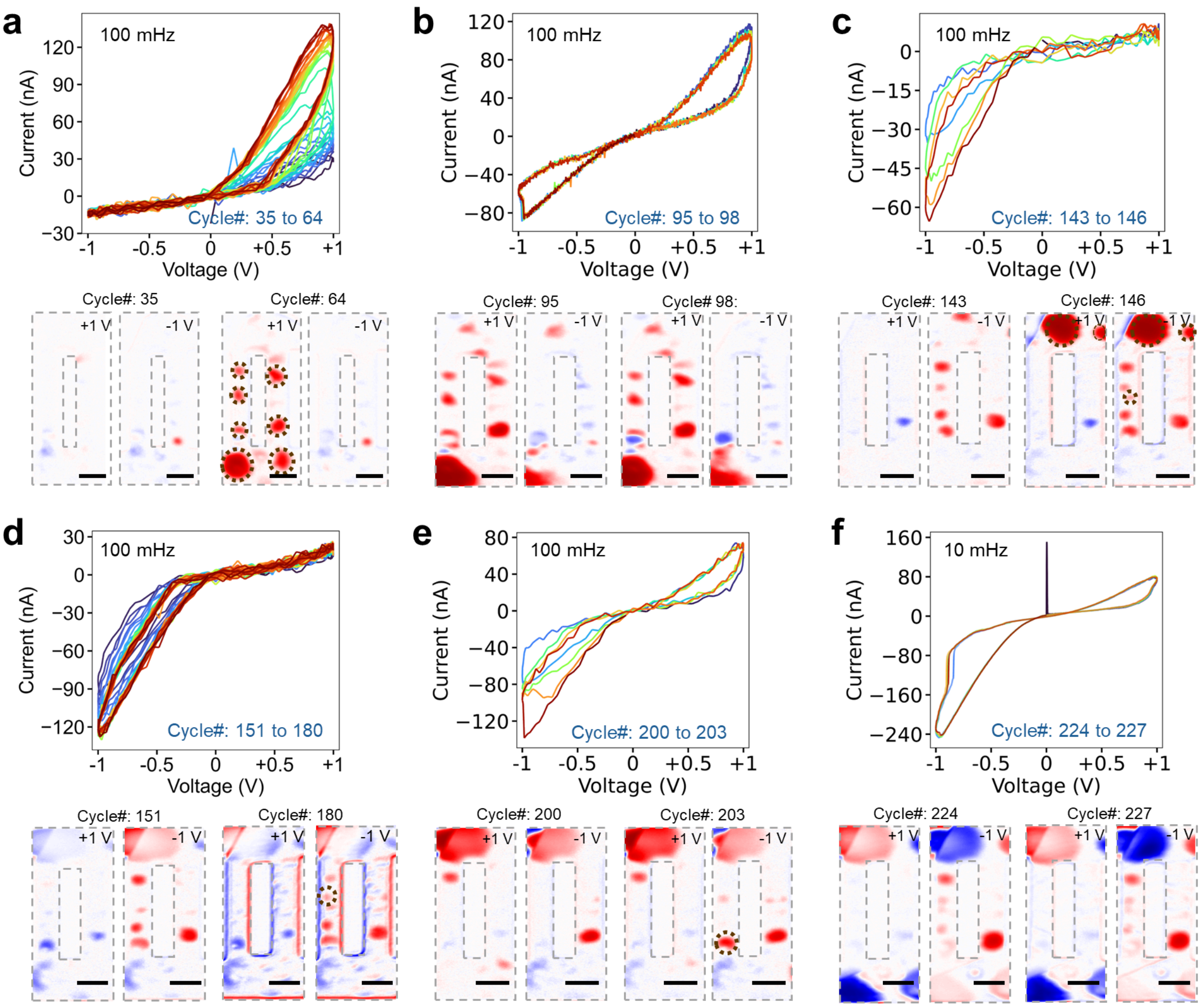

**Figure S20.** Selected I-V cycles of Device 2 alongside VIC images at the corresponding cycle numbers. Cycles are drawn from the full 227 cycle data set in Figure S19, recorded in 1 M KCl at ±1 V and frequencies of 100 mHz (panels a-e) and 10 mHz (panel f). The selected cycles capture the transitions in I-V and the corresponding VIC images and show that these transitions coincide with the appearance of new blisters. The newly formed blisters are indicated by dashed brown circles in VIC images. In panel A, despite the voltage range of ±1 V, the device exhibits a strictly linear off-state response (negative polarity) in the absence of any blister formation consistent with the low voltage response in Figure 2a confirming active structural evolution is required for hysteresis irrespective of the applied voltage. *I-V* traces are coloured blue to red in order of increasing cycle number. Scale bar: 5 µm

## 7.9 Blisters can form at any interface.

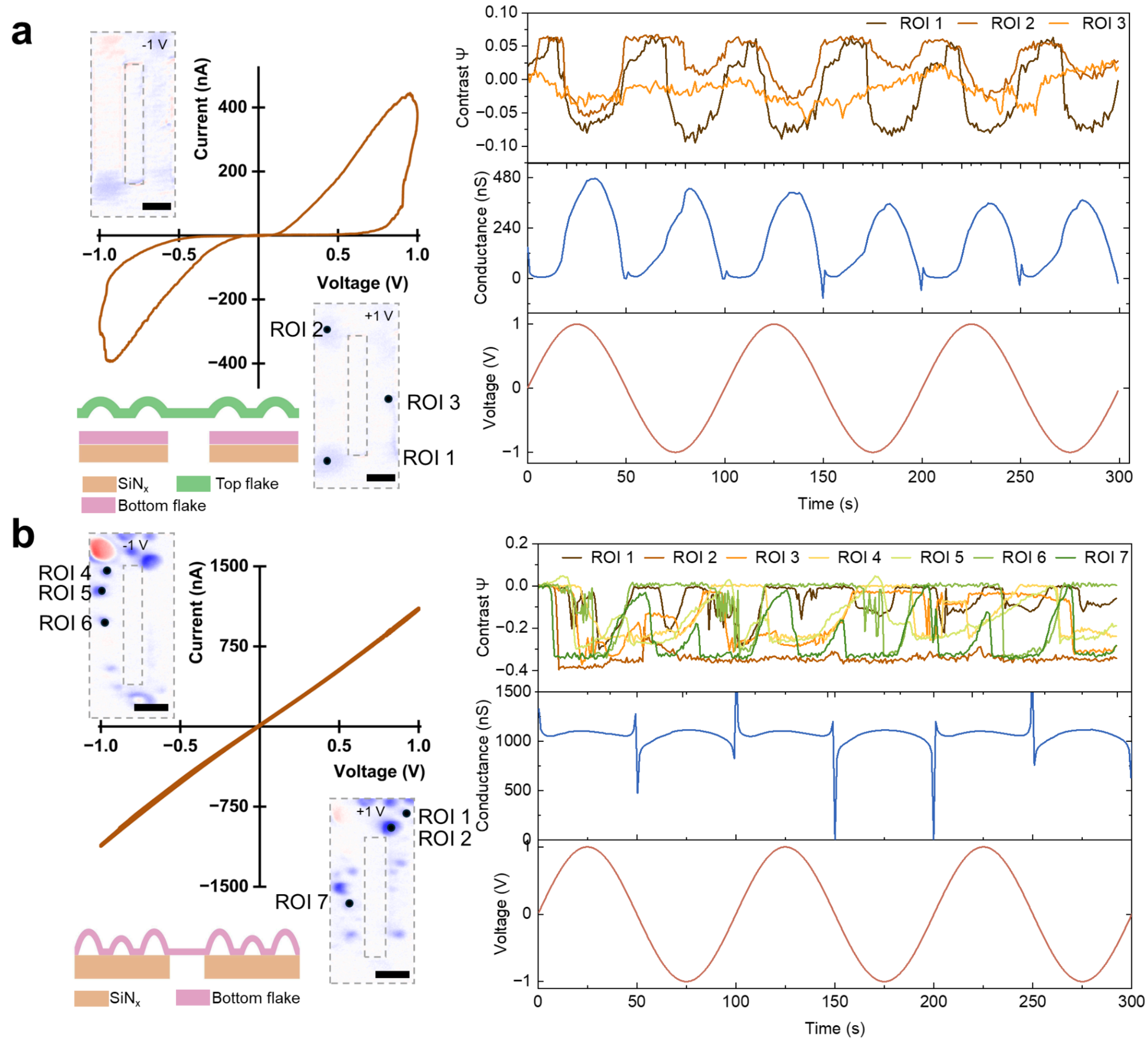

**Figure S21. Blisters can form at any interface of the heterostructure:** *I-V* curves and VIC images and time traces of contrast of selected ROIs, device conductance and applied voltage for two devices measured in 1 M KCl at a frequency $f$ = 10 mHz. (a) Device 9 (graphite bottom): the VIC images show blisters forming above the graphite layer, since the graphite is opaque and any deformation beneath it would not be optically visible. (b) Device 12 (hBN on $SiN_x$): blisters are visible on the bare hBN surface indicating that the delamination can also occur at the bottom interface. Together, panels a and b show that blister nucleation is not confined to the top surface and it can occur at any interface within the heterostructure (above or below). ROIs 1-3 (panel a) and ROIs (panel b) mark the regions used to extract the optical contrast time traces. Scale bar: 5 µm

## 7.10 Effect of salt concentration

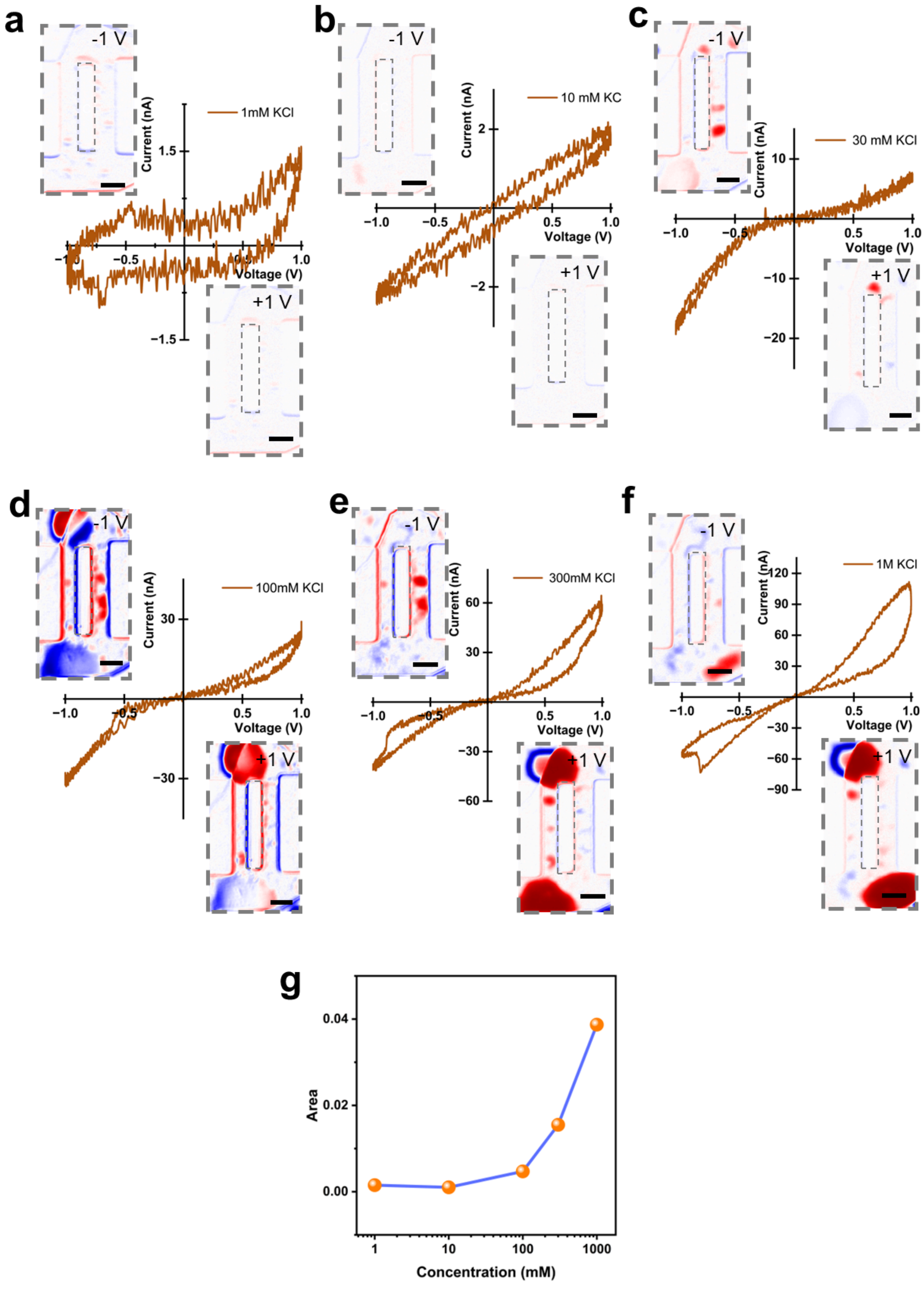

**Figure S22.** (a-f) *I-V* curves and VIC images of device 2 at ±1 V with varying concentration of KCl from 1 mM to 1 M at a frequency *f* = 10 mHz and (g) shows corresponding hysteresis loop area vs concentration plot . At low concentrations (1-30 mM), the *I-V is* nearly linear with 30 mM showing rectification in *I-V.* and VIC images show little contrast. Both the hysteresis loop area and VIC increases above 100 mM with 1 M KCl producing the largest loop area. The concentration dependence in panel g indicates a threshold behaviour with memristive response emerging only once the ionic strength exceeds a critical value. Scale bar: 5 µm

# 8. Voltage-induced contrast statistics and blister spatial distribution across devices

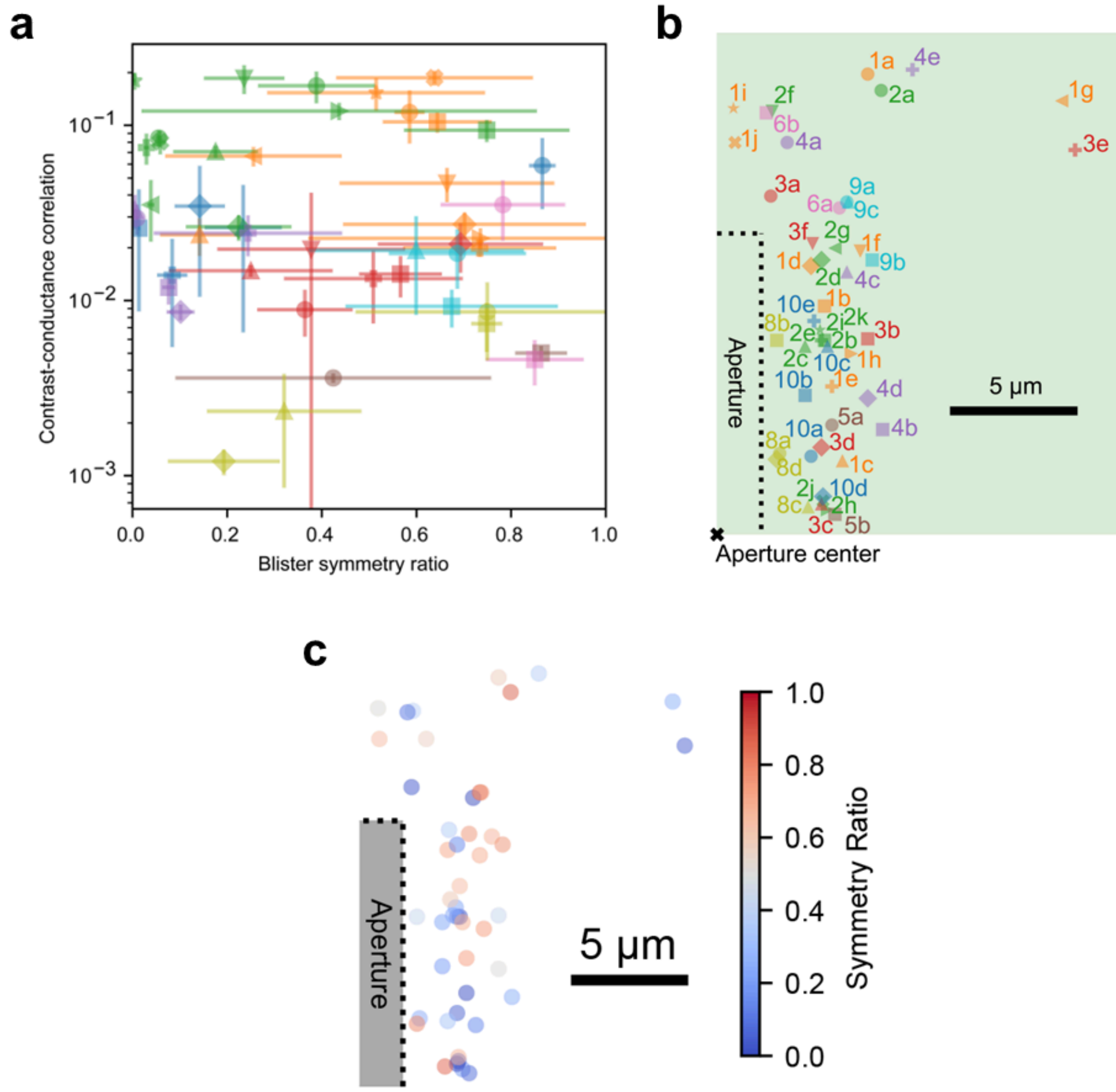

**Figure S23. (a)** Scatter plot of *C* versus *S* for all blisters analyzed in hBN devices (N = 47 blisters from 10 devices) as shown in Fig. 5b with error bars (values obtained from 2 to 4 cycles), (b) Extended spatial distribution of blister positions relative to the aperture centre, overlaid for all devices and (c) Blister positions as (b) with each blister colour coded by its symmetry ratio S. Close to the aperture, both high and low symmetry blisters are found, whereas far from the aperture low-symmetry (unidirectional) blisters dominate. This observation is consistent with the osmotic model for bidirectional blisters, which requires the blister to be located on the ion transport path.

# 9. Theoretical modelling of the observed deformations

## 9.1. Blister modelling and energy barrier

### 9.1.1 Energy functional

We consider a nanochannel of hBN of length $L$ = 5 µm and thickness $t$ ~ 100 nm with strong confinement of $h$ = 5 nm in which we use an electrolyte at ionic concentration of $\rho_s$ = 1 M and apply on it a voltage drop of $\Delta V$ = 1 V. Typical blisters have a height of $h_B$ ~ 10 nm over a radius $R$ ~ 1 µm.

The blister is an elastic deformation modelled by a deflection of height $h_B$, whose energy cost is controlled by the Young's modulus of hBN out-of-plane $Y_\perp \approx 30$ GPa and the in-plane $Y_\parallel \approx 800$ GPa. In practice, the effective Young's modulus intervening in the bending energy of multi-layer hBN decreases with the number of layers: $Y_\perp^{eff} \approx Y_\perp/N$ where $N$ is the number of layers, so that the bending is negligible for the experimental configuration[4]. Moreover, the hBN layer lays on spacers in graphene, with which it has an adhesion energy. A blister which covers several nanochannels needs to break its adhesion over the area of contact with the spacers. However, this area can be reduced by already existing bubbles (introduced for instance, during fabrication) which do not break the tightness of the van der Waals assembly. We denote $\Gamma$ as the adhesion energy per unit surface of hBN on graphene[5]. Overall, the total energy of a hemispheric blister of radius R and height $h_B$ experiencing a pressure $P$ is then given by:

$$\Delta E \approx \alpha\pi R^2\Gamma + (\pi Y_\parallel t / 12R^2)\, h_B^4 - (P\pi R^2/2)\, h_B \qquad (1)$$

The first term is the adhesion energy to break. The second term is the stretching energy of the deformed hBN layer. The last term is the work of the applied pressure.

### 9.1.2 Blister energy threshold

For the blister to form, the first step is to overpass the energy barrier of adhesion. In practice, both $h_B$ and $R$ grow together. However, $h_B(t)$ is a growing function of time which is limited by the stretch energy cost. Minimising the energy over $h_B$ leads to this limit deflation:

$$h_B^0 = [3PR^4 / (2Y_\parallel t)]^{1/3} \qquad (2)$$

As long as $h_B < h_B^0$, the height grows and the energy cost of expanding the blister (increasing R) is decreasing. Hence, the blister will start expanding at some point unless it remains energetically unfavourable when the height $h_B$ has reached its limit value $h_B^0$. It is therefore sufficient to focus on the energetics in the limit case $h_B = h_B^0$.

Thus, the effective energy of the blister reads[6]:

$$\Delta E \approx \pi R^2\, [\Gamma - (3/8)(3R^4 / 2Y_\parallel t)^{1/3}\, P^{4/3}] \qquad (3)$$

We immediately notice that R = 0, corresponding to the absence of blisters, is a (local) minimum of the energy function. Small blisters are energetically forbidden while large blisters become energetically favourable. In practice, the size of the blisters is however limited by the size of the system. The largest possible blisters, which are the most favorable energetically, then have a size $R \sim L \approx 5$ µm. This explains why the experimentally observed blisters of a size comparable to the system while small blisters are not observed. For these blisters to appear their energy must be negative, which provides a threshold on the required pressure:

$$P_c \approx [(2Y_{\parallel}t)^{1/4} (8\Gamma)^{3/4}] / (3R) \qquad (4)$$

Note that this threshold depends on space if the adhesion is not uniform. Even for a single blister, the adhesion might be non-uniform leading to a distribution of thresholds rather than a single value, which would result in the experimental data to a smoothened threshold.

### 9.1.3 Nucleation

We can also study the energy differential when the blister grows:

$$\mathrm{d}\Delta E/\mathrm{d}R \approx 2\pi R\, [\, \Gamma - (7/8)(3R^4 / 2Y_{\parallel}t)^{1/3} P^{4/3}] \qquad (5)$$

which is negative for R large enough. This means that a blister can form only if there is a zone of low adhesion which is large enough. The differential becomes negative for pressure larger than:

$$P_c \approx [(2/3\; Y_{\parallel}t)^{1/4} (8/7\; \Gamma)^{3/4}] / R \qquad (6)$$

which is very similar to Eq. (5).

### 9.1.4 Hysteresis and memory effect

Once the blister is formed, the adhesion energy vanishes and there is no more threshold to cross. Therefore, when the voltage decreases, the blister size decreases smoothly with:

$$h_B = [3PR^4 / (2Y_{\parallel}t)]^{1/3} \qquad (7)$$

When the pressure is below the threshold, the blister remains until the height reaches zero and adhesion is restored. Thus, it keeps memory of the history: whether the threshold has already been crossed or not since the last zero voltage.

### 9.1.5 Dynamical effects

The previous analysis was purely static: we determined that is the state which minimises the energy. In practice however, the evolution towards a new mechanical minimum might take some time. In all practical cases, the inertia of the deformed layer remains negligible. We identify two main effects which might be dynamically limitant: the de-adhesion process and the filling of the blister when growing once the threshold has been reached. The former is known to contain many timescales and is hard to evaluate theoretically[7]. The latter can be estimated roughly.

Once the threshold is reached, the energy landscape pushes the blister deflation towards $h_B$

given by Eq. (7) and the blister grows. Nevertheless, for a nanochannel the hydrodynamic resistance $R_h$ is large. Therefore, ensuring a flow rate requires a strong pressure drop. Assuming for simplicity that the blister has a radius R and no curvature, we estimate the required pressure drop to fill the growing blister as $Q \approx \pi R^2 \partial_t h_B \approx \Delta P / R_h$ . This pressure drop reduces the pressure $P_c$ which pushes the blister to grow and acts as a dissipation for the blister dynamics. In the overdamped regime, we then have the dynamical equation:

$$\pi R^2 \Delta P = \pi^2 R^4 R_h \partial_t h_B = -(\pi Y_{\parallel} t / 3R^2)\, h_B^{\,3} + (P_c \pi R^2/2)$$

The timescale to grow the blister can then be estimated by

$$\tau \sim \frac{3\pi R^6 R_h}{Y_{\parallel} t\, h_B^{\,2}} \sim \frac{36\pi R^6 \eta l}{w h^3\, Y_{\parallel} t\, h_B^{\,2}}$$

where $w$ is the effective width of the channels, $l$ its length and $h$ its height. Here for simplicity we take $w \sim l \sim R$ while $h \approx 1$ nm and $h_B \approx 10$ nm. Thus, we obtain $\tau \approx 10$ s.

Thus, while an exhaustive investigation of the timescales at play in these memristors is beyond the scope of this paper, we provided a first rough estimate for the frequency cutoff of the memristor around 100 mHz. We expect the memristive effect to decrease at larger frequencies as a blister has not enough time to grow.

## 9.2. Bidirectional memristors: Polarisation concentration-induced osmotic pressure

### 9.2.1 Electroneutrality, resistivity and concentration profile

Since $L \gg \lambda_D$ we assume electroneutrality. Indeed, the electroneutrality will be recovered over the scale of the Debye length thanks to the electric field created by the charge distribution. Thus,

$$c_+ - c_- = -2\Sigma/h \qquad (8)$$

According to Nernst-Planck's law, the total flux reads:

$$J_{tot} = J_+ + J_- = -D\partial_x c_{tot} - (2\mu\Sigma/h)\, E(x) \qquad (9)$$

and must be uniform. Here, $c_{tot}=c_+ + c_-$ is the total ion concentration including the surface countercharges. The electric current, including the surface contribution, reads

$$I = J_+ - J_- = c_{tot}(x)\, E(x)$$

and must also be uniform. Because the electroneutrality may be locally broken, the electric field may depend on space. This local Ohm's law simply states that the local polarisability accommodates the local electric field to the local ionic resistivity $\rho_i \propto 1/c_{tot}$. Thus,

$$E(x) = -\Delta V/(c_{tot} \int dx/c_{tot}) \quad (10)$$

For a large Dukhin number $Du = |\Sigma|/ehc_0$, the channel becomes selective and the total flux $J_{tot}$ becomes small compared to the electric current I. In this limit, there is a strong polarisation concentration effect and an equilibrium concentration profile. This limit is valid for low salt concentration. For smaller Dukhin numbers the selectivity is partial and the total flux is not negligible. In the experiment, the Dukhin number ranges from intermediate to large values. Nevertheless, we will assume a perfect selectivity in the following in order to allow the model to be tractable and to provide qualitative insights on the physics at play, keeping in mind that we overestimate the effect at large salt concentration. With this assumption:

$$D\partial_x c_{tot} = (2\mu\Sigma/h)(1/c_{tot})\,(\Delta V/\int dx/c_{tot}) \quad (11)$$

For simplicity, we denote $\xi = e\Delta V/k_BT$ and $R = \int dx/c_{tot}$ , and we adimensionalise $c_{tot}$ by $2c_0$ and x by $L$. Then,

$$c_{tot}(x) = [c_{tot}(0)^2 + (Du\,\xi / 2R)\,x]^{1/2} \quad (12)$$

Also,

$$c_{tot}(1) = [c_{tot}(0)^2 + Du\,\xi / 2R]^{1/2} \quad (13)$$

At the edges the concentrations are not bulk yet due to the concentration polarisation effect. Yet, they are balanced by the diffusion flux between the entrance of the channel and the far away reservoir at bulk concentration. Using symmetric reservoirs and because we must have equal fluxes at each edges, we then obtain:

$$c_{tot}(0) + c_{tot}(1) = 2 + 2Du \quad (14)$$

Thus,

$$c_{tot}(0)^2 + Du\,\xi / 2R = [2 + 2Du - c_{tot}(0)]^2 \quad (15)$$

That is,

$$c_{tot}(0) = 1 + Du - Du\,\xi / [8R(1+Du)] \quad (16)$$

$$c_{tot}(1) = 1 + Du\,\xi / [8R(1+Du)] \quad (17)$$

The resistivity denominator reads:

$$R = \int dx/c_{tot} = 4R/Du\,\xi\,[c_{tot}(1) - c_{tot}(0)] = 1/(1+Du) \quad (18)$$

in which the Dukhin number corresponds to the surface transport contribution to the conductance. Thus,

$$c_{tot}(x) = [\,(1 - Du\,\xi/8)^2 + (1+Du)\,Du\,\xi/2\,x]^{1/2} \quad (19)$$

### 9.2.2 Osmotic pressure

Finally, the average density reads:

$$\langle c_{tot} \rangle = 4\,(c_{tot}(1)^3 - c_{tot}(0)^3)/[3Du\,\xi(1+Du)] \qquad (20)$$

that is

$$\langle c_{tot} \rangle = 4/3\, Du\, \xi(1+Du)[3Du\xi(1+Du)^2/4+2Du^3\, \xi^3/8^3] = 1+Du+Du^2\, \xi^2/[192(1+Du)] \qquad (21)$$

meaning that we have an over-concentration inside the channel, corresponding to an average osmotic pressure of:

$$\Delta\Pi = 2c_0 T\, (\langle c_{tot} \rangle - 1 - Du) = (c_0 T Du^2 \Delta V^2)/(96(1+Du)T^2) = \Sigma^2 \Delta V^2/[96(1+Du)\, h^2\, c_0\, k_B T] \qquad (22)$$

where we have reintroduced the dimensions in the last expressions. Note that the pressure is positive independently of the sign of $\Delta V$.

Using Eq. (7), this model predicts that the height or contrast of such a blister in its receding dynamics scales with $\Delta V^{2/3}$, which is consistent with experimental data (see Fig. S24a).

## 9.3. Unidirectional memristors: Electrostatic pressure

### 9.3.1 Capacitive charging

First of all, let us check how the capacitive effect compares with the surface charge. We note that the top layer has capacitance around:

$$C_0 \approx \varepsilon_0\, \varepsilon_{hBN} / t \approx 0.5\ \mathrm{mF/m^2} \qquad (23)$$

The capacitive charging due to the voltage drop can then be estimated as:

$$\Delta\Sigma = C_0\, \Delta V \approx 0.5\ \mathrm{mC/m^2} \qquad (24)$$

which is smaller than the typical surface charge on hBN. Thus, we can neglect any effect related to capacitive charging.

### 9.3.2 Debye double-layer

The Debye double-layer adds a capacitance $C_D$ in parallel to the hBN layer on each side, thus reducing the voltage drop experienced by the solid:

$$\Delta V_{eff} \approx \Delta V / (1 + 4C_0/C_D) \qquad (25)$$

We can estimate the ratio of capacitance as:

$$C_0/C_D \approx \lambda_D/t \times (\varepsilon_{hBN}/\varepsilon_w) \approx 10^{-3} \qquad (26)$$

so that the Debye's capacitance is much larger and then short-circuited by the solid's capacitance. Therefore, we can neglect the capacitive contribution from the Debye double-layer.

### 9.3.3 Electrostatic pressure due to surface charge

The surface charge will experience the local electric field and then induce a pressure force on the solid. Since the surface charges we consider are in the first layer close to the interface, they are screened by water in the parallel direction but mostly not in the normal direction[8]. Therefore, the normal contribution of the electric field at the surface mostly fully drags the surface charges. The situation is quite different for the counter charges which are further from the interface and therefore screened by water in every direction. These countercharges thus experience a much smaller local electric field in the normal direction. This justifies focusing only on the effect on the surface charges of the normal component of the electric field. Due to the geometry of the system and given that we can neglect of the capacitive effect of the Debye double-layer, the normal electric field going throughout the top layer is estimated close to the center as:

$$E_{\perp} \approx \Delta V / (\varepsilon^{\perp}_{s}\, t) \tag{27}$$

where $\varepsilon^{\perp}_{s} \approx 2.1$ according to ref 8. Thus, the force applied on the surface charge can be estimated as:

$$P_{elec} = 2\Sigma E_{\perp} \approx 2\Sigma\, \Delta V / (\varepsilon^{\perp}_{s}\, t) \tag{28}$$

which should be enough to reach the threshold to form a blister.

Using Eq. (7), this model predicts that the height or contrast of such a blister in its receding dynamics scales with $\Delta V^{1/3}$, which is consistent with experimental data (see Fig. S24a).

## 9.4. Dependence on salt concentration

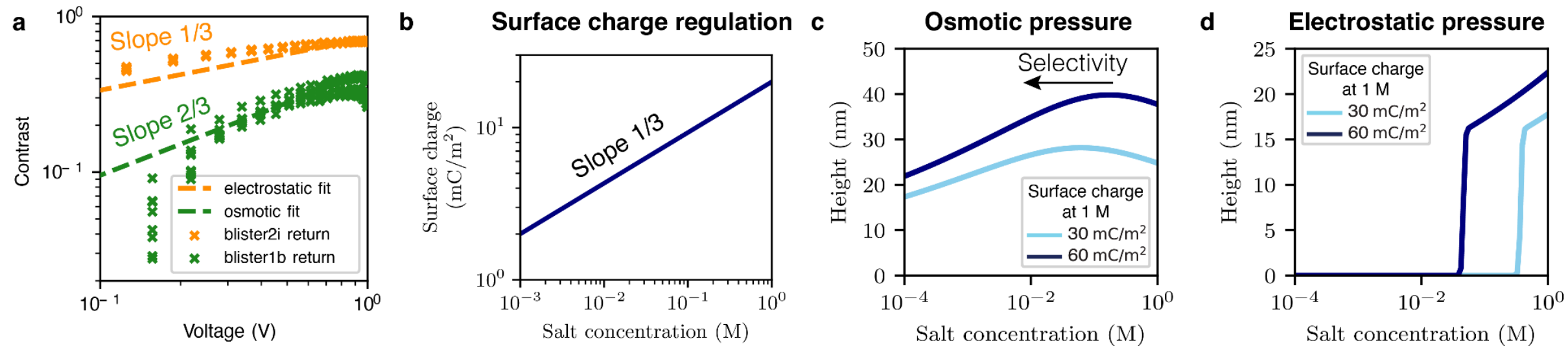

**Figure S24. a,** Scaling of the contrast (or equivalently height) in the receding dynamics as a function of voltage for both models and for experimental data of model blisters (1b for the bidirectional memristor and 2i for the unidirectional memristor). **b,** Model used for the surface charge regulation: $\Sigma \sim c_0^{1/3}$. **c,** Prediction of maximum blister height as a function of salt concentration in the model of osmotic pressure. **d,** Prediction of maximum blister height as a function of salt concentration in the model of electrostatic pressure.

Due to surface charge regulation, the surface charge is expected to depend on the salt concentration[9]. While this dependence adds complexity, we assume a standard scaling $\Sigma \sim c_o^{1/3}$ (see Fig. S24b). Assuming such a scaling, we can study the maximum blister height at 1 V as a function of the salt concentration: the results are given in S24c-d. We recover the experimental insight that the effect disappears at small concentrations (Supplementary Section 9.2). For the electrostatic pressure (Fig. S24d), at lower salt concentration, the surface charge is smaller and no longer reaches the adhesion threshold between 50 and 500 mM.

For the osmotic pressure (Fig. S24c), the effect is more subtle because the selectivity also depends on the salt concentration: at large concentration there is a small Dukhin number and then a low selectivity. For our model to be tractable, we have assumed a strong polarisation effect in order to neglect the total salt current. This approximation becomes invalid at low selectivity, that is above 200 mM ($Du \sim 1$). Below this salt concentration, the osmotic pressure is predicted to scale with the surface charge and then to increase with the salt concentration. For larger salt concentrations, our model is no longer valid but we do expect a saturation. Thus, we can expect the extrapolation of our model up to 1 M to remain qualitatively acceptable.

## 10. Supplementary discussion: evidence that electromechanical hysteresis overrides purely ionic contributions to memory

In this work, our operando optical interferometry directly measures nanoscale structural deformations rather than local ion concentrations. Based on these observations and in particular, the symmetry of the voltage response, we infer that ionic concentration polarization (CP) likely provides the osmotic pressure responsible for bidirectional blister actuation. Nevertheless, we have strong evidence that purely ionic effects cannot account for the memristive hysteresis observed in these 2D nanochannels. Attributing the observed conductance memory solely to ionic microstates leads to fundamental physical contradictions. Instead, our measurements and theoretical boundary calculations establish that macroscopic structural deformation acts as the primary state variable governing the memory effect. This conclusion is supported by three independent lines of direct evidence:

**1. Static vs. dynamic geometry: the absence of memory without deformation**
The necessity of mechanical deformation is evidenced by the absence of memory when the channel geometry remains static. In highly confined regimes where CP is active, the device response remains linear if the van der Waals slit does not actively deform (Main text Fig. 2a). This is true for higher voltage ranges where devices show no active blister deformation (OFF-state conductance at negative voltages in Fig. 4d, fully linear IV curves in Fig. S9b, negative and positive parts of the IV curves in Fig. S20a and S20c, respectively). Hysteresis only emerges in precise spatiotemporal synchrony with dynamic blister growth and relaxation. Furthermore, control devices operating in regimes of negligible ionic selectivity (e.g., $h \approx 50$ nm slits) show clear blister nucleation driven by electrostatic forces without exhibiting macroscopic *I-V* hysteresis (Fig. S11c), demonstrating that blisters can form independently of CP.

**2. Baseline delamination and conductance limits**
To quantify the geometric contribution to the observed memory, we calculate the theoretical conductance of a perfectly flat, rigid van der Waals nanoslit. For a standard device geometry ($N$=60, $L \approx 4$ µm, $w$=100 nm, and $h$=0.7 nm) filled with 1 M KCl (conductivity $\sigma \approx 10$ S/m), the baseline conductance is defined as:

$G_{flat}$=$N \cdot \sigma \cdot w \cdot h/L$ ≈ 10.5 nS. The experimentally measured baseline of ~40 nS at 0 V confirms that the channel is inherently non-rigid and already partially delaminated prior to out-of-plane biasing.

**3. Physical limits of ionic memory**

During dynamic operation, the memristive switching exhibits discrete conductance jumps. When a characteristic dynamic blister forms (e.g., with an estimated height $h_B$≈10 nm), its structural contribution to the channel cross-section can be approximated as $G_{blister} \approx \sigma \cdot h_B$≈100 nS. This purely geometric deformation observation perfectly accounts for the measured ~100 nS discrete conductance jumps.

Conversely, attributing these massive conductance modulations solely to ionic changes is physically impossible. The devices routinely exhibit ON/OFF conductance ratios of approximately 50 (Main Text, Fig. 3c-d). Generating a 50-fold conductance increase in a rigid 0.7 nm channel via ionic effects alone would require the local KCl concentration to surge from 1 M to 50 M. Because the theoretical aqueous solubility limit of KCl is ~4.8 M, such a concentration is unattainable. Therefore, while ionic pressure triggers the phenomenon, the macroscopic mechanical deformation of the compliant channel walls is the necessary physical state variable that enables the memristive hysteresis.